\documentstyle[times,epsfig]{mn}

\newcommand{\bez}{\begin{eqnarray*}}
\newcommand{\eez}{\end{eqnarray*}}
\newcommand{\be}{\begin{equation}}
\newcommand{\ee}{\end{equation}}
\newcommand{\beq}{\begin{eqnarray}}
\newcommand{\eeq}{\end{eqnarray}}
\newcommand{\bc}{\begin{center}}
\newcommand{\ec}{\end{center}}

\newbox\grsign \setbox\grsign=\hbox{$>$} \newdimen\grdimen \grdimen=\ht\grsign
\newbox\simlessbox \newbox\simgreatbox \newbox\simpropbox
\setbox\simgreatbox=\hbox{\raise.5ex\hbox{$>$}\llap
     {\lower.5ex\hbox{$\sim$}}}\ht1=\grdimen\dp1=0pt
\setbox\simlessbox=\hbox{\raise.5ex\hbox{$<$}\llap
     {\lower.5ex\hbox{$\sim$}}}\ht2=\grdimen\dp2=0pt
\setbox\simpropbox=\hbox{\raise.5ex\hbox{$\propto$}\llap
     {\lower.5ex\hbox{$\sim$}}}\ht2=\grdimen\dp2=0pt
\def\simgt{\mathrel{\copy\simgreatbox}}
\def\simlt{\mathrel{\copy\simlessbox}}

\def\intl{\int\limits}

\def\tT{\tau_{\rm T}}
\def\sigmat{\sigma_{\rm T}}
\def\sigmaph{\sigma_{\rm ph}}

\def\Hlam{H_{\lambda}}

\def\cosi{\cos i}
\def\sini{\sin i}
\def\d{{\rm d}}
\def\Rin{r_{\rm in}} 
\def\Rout{r_{\rm out}} 

\def\Rg{R_{\rm g}} 
\def\rmin{r_{\rm m}}
\def\rmax{r_{\rm M}}
\def\pmin{p_{\rm m}}
\def\pmax{p_{\rm M}}
\def\thmax{\theta_{\rm M}}  
\def\thmin{\theta_{\rm m}}  
\def\tmax{t_{\max}}  
\def\tmin{t_{\min}}  
\def\calR{{\cal R}}

\def\hT{\hat{T}}
\def\hI{\hat{L}}
\def\hC{\hat{C}}
\def\hID{\hat{L}^{\rm D}}

\def\ID{L^{\rm D}}

\def\IDE{L^{\rm D}_E}

\def\IE{L_E}
\def\aE{a_{E}}
\def\alam{a_{E}}
\def\rhoe{\rho_E}
\def\aeff{a_{\rm eff}}
\def\RE{A_E}
\def\RDE{A^{\rm D}_E}
\def\RT{A^{\rm T}}
\def\CD{C^{\rm D}}

\def\vp{\mbox{\boldmath $p$}}
\def\vn{\mbox{\boldmath $n$}}
\def\vd{\mbox{\boldmath $d$}}
\def\vo{\mbox{\boldmath $o$}}
\def\vh{\mbox{\boldmath $h$}}

\def\xte{{\it RXTE}}

\begin{document}

\title[X-ray reverberation  in flared accretion  discs] 
{Impact of reverberation in flared accretion discs on 
temporal characteristics of X-ray binaries}


\author[Juri Poutanen]
{\parbox[]{6.8in} {Juri~Poutanen$^{\star}$}\\
Astronomy Division, P.O. Box 3000, 90014 University of Oulu,
Finland\\
Stockholm Observatory, 10691 Stockholm, Sweden
}

\date{Accepted, Received}

\maketitle


\begin{abstract}
Observations  suggest that accretion  discs in many X-ray binaries are likely flared.
An outer  edge of the disc  intercepts  radiation  from the  central  X-ray
source.  Part of that radiation is absorbed and reemitted in the optical/UV
spectral  ranges.  However, a large fraction of that radiation is reflected
and appears in the broad-band X-ray spectrum as a Compton  reflection bump.
This radiation is delayed and variability is somewhat smeared compared with
the intrinsic X-ray radiation.  We compute response  functions for flat and
flared accretion discs and for isotropic and anisotropic  X-ray sources.  A
simple  approximation for the response function which is valid in the broad
range of the disc shapes and  inclinations,  inner and outer radii, and the
plasma bulk  velocity  is  proposed.  We also study the impact of the X-ray
reprocessing  on temporal  characteristics  of X-ray  binaries  such as the
power  spectral  density,  auto-  and   cross-correlation   functions,  and
time/phase lags.  We propose a
reprocessing  model which  explains  the  secondary  peaks in the phase lag
Fourier spectra  observed in Cyg X-1 and other Galactic black hole sources.
The  position  of the  peaks  could be used to  determine  the  size of the
accretion disc.
\end{abstract}



\section{Introduction}

\footnotetext{$^\star$ E-mail: juri.poutanen@oulu.fi}

The X-ray spectra of  radio-quiet active  galactic  nuclei  (AGNs) and 
X-ray  binaries can often be  decomposed into a power-law like continuum 
with a cutoff at 100 keV (Zdziarski  1999), 
soft black body  emission from the accretion disc and/or neutron star
surface,  and the Compton  reflection  continuum with the associated
iron  emission  line at $\sim 6.4$ keV (see e.g.  George \& Fabian 1991;
Nandra \& Pounds 1994;    Reynolds  1999).  The  reflection
features  are  produced in a rather cold  neutral  material  which is often
identified  with an accretion  disc.  Correlation  between the amplitude of
reflection, $R$, and the photon spectral index, $\Gamma$, in X-ray binaries
and AGNs  (Zdziarski,  Lubi\'nski  \& Smith  1999;  Gilfanov,  Churazov  \&
Revnivtsev  2000a)  implies that a relatively large  fraction of reflection
originates  close  to the  X-ray  emitting  region.  In the  case of  X-ray
binaries, some fraction of reflected photons, however, can also come from a
companion  star (Basko,  Titarchuk \& Sunyaev  1974; Done \& \.Zycki  1999;
Vikhlinin  1999) or an outer  part of the  accretion  disc.  In the case of
AGNs,  the  delayed   reflection  from  a  distant  molecular  torus  (e.g.
Ghisellini,  Haardt \& Matt 1994) can be clearly  observed when the central
X-ray source turns off (Guainazzi et al.  1998).

The  reflected  radiation  is  necessarily  delayed  relative to the direct
radiation  from the  X-ray  source.  Studies  of these  delays  can help to
determine  the geometry of the  accretion  disc, the  position of the X-ray
source, and the distance to the  companion  (Vikhlinin  1999).  A number of
papers were devoted to studies of the response of the Fe line profile to an
X-ray flare in the  vicinity of a black hole (e.g.  Reynolds  et al.  1999;
Hartnoll  \&  Blackman  2000).  Observations  of  AGNs  with  future  X-ray
missions such as  Constellation-X  may open a  possibility  of  determining
directly from the Fe line profile  temporal  evolution, the geometry of the
X-ray  emitting  region and the mass and spin of black holes (see  Reynolds
1999;  Fabian  et al.  2000 for  reviews).  However,  in order  to do that,
observation  of  a  bright  individual  flare  is  needed.  Otherwise,  the
interpretation  of the  line  profile  produced  by many  flares  would  be
complicated.  Such  studies are  impossible  in the case of X-ray  binaries
even with future  instruments since the photon flux per light crossing time
of one gravitational radius is thousands times smaller than in AGNs.  As an
alternative  to studying the line profiles one can use the whole  available
statistics  at  different  energy  bands and analyse  the  response  of the
reflected {\it continuum} radiation to a varying X-ray source.

All the temporal studies mentioned above assumed that the accretion disc is
flat.  In contrast, a number of observations  indicate that accretion discs
are flared, i.e.  geometrically  thick at the outer edge (see  Verbunt 1999
for a  recent  review).  There  observations  include  obscurations  of the
central  X-ray  source  (e.g.  White  \& Holt  1982)  and  the  variability
properties  in the  UV/optical/infrared  spectral  band in  low-mass  X-ray
binaries  (e.g.  Mason \& Cordova  1982;  Vrtilek et al.  1990), the delays
between  the  optical/UV  emission  and the  X-rays  observed  in the X-ray
bursters  (Tr\"umper  et al.  1985)  and in the  super-luminal  black  hole
source  GRO~J1655-40  (Hynes et al.  1998).  The inferred  height-to-radius
ratio,  $H/\Rout$,  at the outer edge of the disc varies  from 0.15 to 0.5.
This is much larger than  predicted by the standard  accretion  disc theory
(Shakura \& Sunyaev 1973).

In this  paper, we study the impact of the X-ray  reflection  from flat and
flared  accretion discs on temporal  characteristics  of the X-ray binaries
and AGNs.  In  \S~\ref{sect:resp},  we present the  general  formalism  for
computation  of the response  functions and introduce  approximations  that
allow one to consider a linear  reflection  response.  Formulae  describing
the impact of reflection on the temporal  characteristics  of the composite
(direct  and  reflected)   signal  are  presented  in   \S~\ref{sect:temp}.
Response  functions for the flat and flared discs and the  resulting  phase
lags are presented in  \S~\ref{sect:res}.  Comparison with observations and
discussion is given in  \S~\ref{sect:disc}.  Conclusions  are  presented in
\S~\ref{sect:concl}.

\section{Response from the Accretion Disc} \label{sect:resp}

\subsection{General formulation}

Let  us  consider  an  axially-symmetric  accretion  disc  and  assume  for
simplicity  that the height of the disc surface above the equatorial  plane
is given by a power-law relation
\be  \label{eq:z}
z(r)=H (r/\Rout)^{\alpha}  ,
\ee
where $H$ is the  maximum  disc  height,  $\Rout$ is the disc  radius,  and
$\alpha\ge1$.  The inner  radius  of the disc is  $\Rin$.  Throughout  this
paper,  distances  will be measured  in units of  $\Rg\equiv  2GM/c^2$  and
time--in units of $\Rg/c$.  For all the examples  considered below, we take
$M=10{\rm  M}_{\odot}$,  but   the  results  can be easily  scaled to any
  mass.

\begin{figure}
\centerline{\epsfig{file=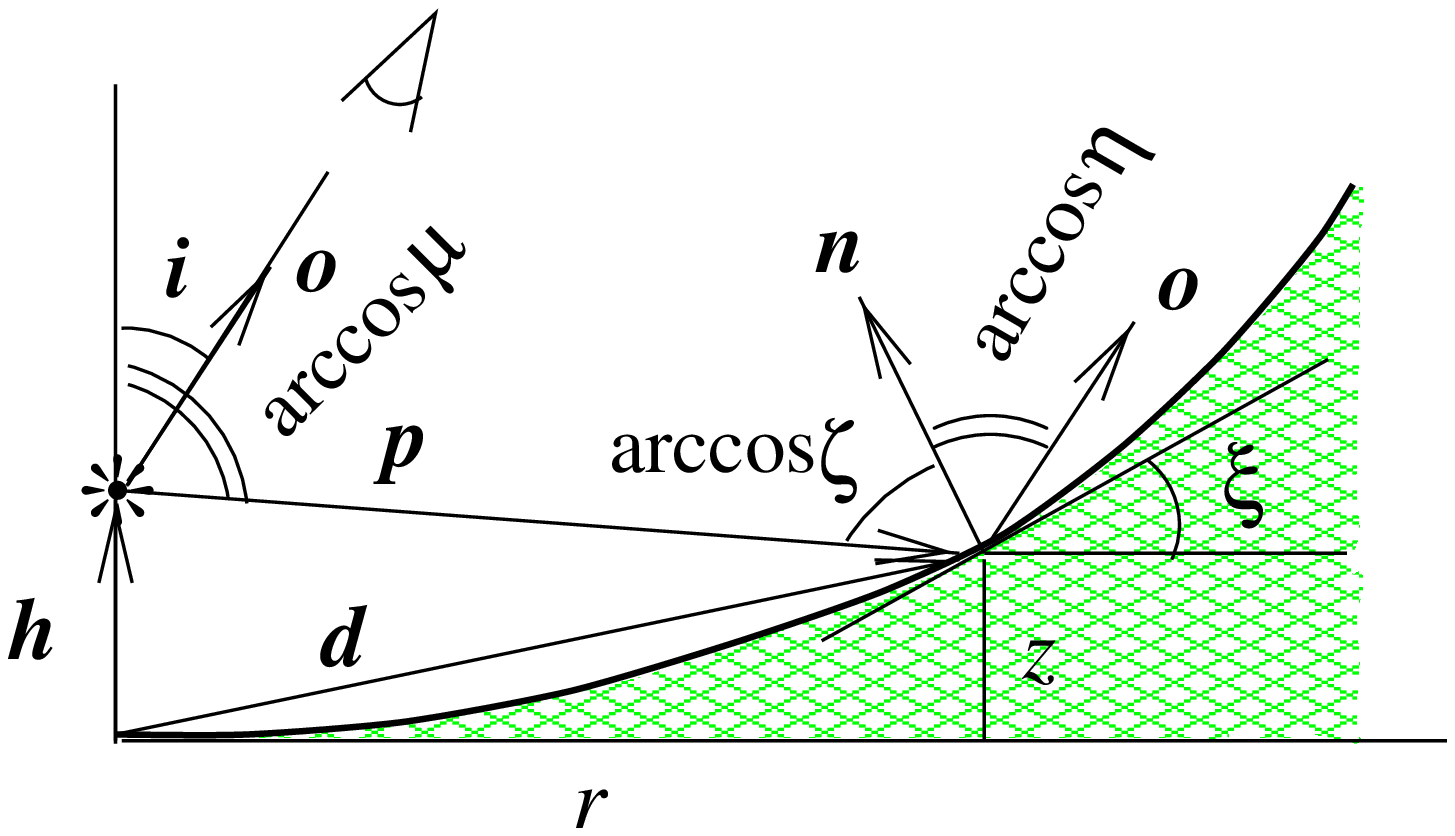,width=7.5cm}}
\caption{Geometry of the problem. 
Note that vector $\vo$ does not generally lie in the plane defined by 
other vectors.}
\label{fig:geom}
\end{figure}

Let us introduce  the  Cartesian  coordinate  system with the origin at the
disc center.  The $x$- and $y$-axes lie in the central  disc plane, and the
line of sight  towards the observer is in the $x-z$ plane in the  direction
$\vo=(\sini,0,\cosi)$,  where $i$ is the  inclination.  The geometry of the
problem is  presented  in  Fig.~\ref{fig:geom}.  An X-ray  point  source is
situated at the disc axis at some  height $h$ from the disc plane, i.e.  at
the position  $\vh=(0,0,h)$.  Let us assume that the X-ray  source  angular
distribution  does not depend on time and  photon  energy.  The   
direct (marked with a superscript $D$) monochromatic luminosity 
(per unit solid angle) is then
\be
L^{\rm D}_E(\mu, t)= \ID_{E}(t) \frac{\Omega(\mu)}{4\pi} .
\ee  
The total emitted luminosity is then 
\be 
L^{\rm D}(t)=\int \d E  \intl_{-1}^{1} \d \mu 2\pi L^{\rm D}_E(\mu, t) .
\ee
Function  $\Omega(\mu)$  gives the  dependence  of the  incident  (direct)
radiation   field  on  zenith  angle   $\arccos\mu$,   and  is  normalised
$\frac{1}{2}\int  \Omega(\mu)\d  \mu =1$, i.e.  for an isotropic  source,
$\Omega(\mu)=1$.

An   element   of   the   disc   surface   has   coordinates    $\vd=(r\cos
\phi,r\sin\phi,z)$,  where the azimuth  $\phi$ is measured  from the $x$-axis.
Photons reflected from that element are delayed by
\be \label{eq:delay}
\Delta t(r,\phi)=p-\vp\cdot\vo=p-r\sini\cos\phi-(z-h)\cosi
\ee
relative   to   the   photons   reaching   the   observer   directly.  Here
$\vp=\vd-\vh$ and $p=\sqrt{r^2+(z-h)^2}$   is  the  distance   between  the 
source  and  the reflection point.

The temporal evolution of the  radiation reflected towards the
observer can be described by the following relation:
\be \label{eq:Lrefl}
L^{\rm refl}_E(\cosi,t)=  \intl_{\rmin}^{\rmax} r \d r  \intl_0^{2\pi}
F_E(\eta,\zeta,r,\phi,t) \d \phi,
\ee
where $\rmin, \rmax$ are  the innermost and outermost disc radii intersecting 
the isodelay paraboloid (see \S~\ref{sec:intazi} and Appendix \ref{sec:app3}) 
and 
\beq\label{eq:ferefl}
F_E(\eta,\zeta,r,\phi,t)&=&
\frac{\Omega(\mu)}{4\pi p^2} \frac{\zeta}{\cos\xi} \frac{2 \eta}{2\pi} 
\nonumber\\
&\times& \intl_{E}^{\infty} \calR(r;E,\eta;E',\zeta) \ID_{E'}(t-\Delta t) \d E' 
\eeq 
is the monochromatic flux of radiation reflected in a unit solid angle from
a surface  element that gives a unit area when  projected  onto the central
plane (note  factor  $\cos\xi=1/\sqrt{1+z'^2}$  in the  denominator,  where
$z'=\partial   z/\partial  r$).  The  reflection   Green  function  $\calR$
describes the processes of Compton  reflection,  photoelectric  absorption,
and fluorescent  line emission (see e.g. Poutanen,  Nagendra \& Svensson  1996).
It depends on the angles  $\arccos\zeta$,  $\arccos\eta$ between the normal
$\vn$ to the disc surface at a reflection  point and the  directions of the
incoming  and  outgoing  photons,  respectively,  as well as on the  photon
energies.  Here
\beq \label{eq:zeta}
\zeta &=& -\vn\cdot\vp/p = [r\sin\xi-(z-h)\cos\xi]/p , \nonumber \\
\eta  &=& \vn\cdot\vo   = -\sini \sin \xi \cos\phi+\cos\xi\cosi , \\
\mu  &=& (z-h)/p . \nonumber
\eeq
In principle,  there exists a weak  dependence of $\calR$ on the difference
in the azimuth of incoming and outgoing photons, which we neglect.  This is
equivalent to assuming isotropic scattering.
The reflection Green function can also be a function of radius, for example,
if the ionisation state of the disc changes with radius.

It is  worth  noticing  that  the  evolution  of  the  reflected  radiation
(eq.~\ref{eq:Lrefl})   depends  not  only  on  time,  photon   energy,  and
inclination  angle,  but  also  on the  shape  of  the  intrinsic  spectrum
$\ID_E(t)$.  Thus, in general,  temporal  characteristics  of the reflected
radiation at a given photon energy $E$ depend on behaviour of the intrinsic
spectrum  at {\it all}  energies  above $E$.  This means that there is {\it
no} linear response between the intrinsic and reflected signals at a single
energy  $E$.  In a number of  physically  realistic  situations  considered
below the  response  is,  however,  very close to linear.  That helps us to
solve temporal problems in a simpler way.

\subsection{Constant intrinsic spectrum} \label{sec:cosp}

If the spectrum of the intrinsic X-ray radiation  $\ID_E(t)$  does not vary
in time while the normalisation varies,
\be
\IDE(t)=P(t)\ID(E) ,
\ee
equation  (\ref{eq:ferefl})  can be simplified:
\be\label{eq:feconst}
F_E(\eta,\zeta,r,\phi,t)= \IDE(t-\Delta t) 
\frac{\Omega(\mu)}{4\pi p^2} \frac{\zeta}{\cos\xi}
\frac{2 \eta}{2\pi} \rhoe(r,\eta,\zeta) , 
\ee 
where we introduced the angle- and energy-dependent  albedo function 
\be \label{eq:rhoe}
\rhoe(r,\eta,\zeta) = \intl_{E}^{\infty} \calR(r;E,\eta;E',\zeta) \ID(E') \d E' 
/ \ID(E). 
\ee
The temporal and all other variables are separated.  The temporal evolution
of the  reflected  radiation  at  energy  $E$ and  inclination  $i$  can be
represented  as a convolution  of the direct  luminosity  at the same angle $i$
with the energy- and inclination-dependent response (transfer) function:
\be \label{eq:lrconv}
L^{\rm refl}_{E}(\cosi,t)= \intl_{-\infty}^{t} 
T_E(t-t')  L^{\rm D}_{E}(\cosi,t') \d t',
\ee
\beq \label{eq:resp}
T_E(t) &=&  
\int r \d r \intl_0^{2\pi} \delta(t-\Delta t) \d \phi  \nonumber \\
&\times& 
\frac{\zeta}{p^2\cos\xi} \frac{\Omega(\mu)}{\Omega(\cosi)}  
\frac{2 \eta}{2\pi} \rhoe(r,\eta,\zeta)  .
\eeq
The total observed luminosity is thus presented as
a sum of the direct and reflected luminosities:
\be \label{eq:ltot}
L_{E}(\cosi,t)=L^{\rm D}_{E}(\cosi,t)
+ \intl_{-\infty}^{t}\!\! T_E(t-t') L^{\rm D}_{E}(\cosi,t')  \d t'  .
\ee

\subsection{Thomson approximation} \label{sec:thom}

X-ray   observations  with  high  temporal   resolution  and  large  photon
statistics exist only at moderate energies ($E\simlt 20$ keV), for which in
the first approximation Compton recoil can be neglected, and the reflection
Green function then takes the form
\be \label{eq:thomson}
\calR(r;E,\zeta;E',\eta)= \rhoe(r,\eta,\zeta) \delta(E-E') .
\ee
In the Thomson  approximation  (\ref{eq:thomson}),  the energy  integral in
equation  (\ref{eq:ferefl}) can be trivially taken.  It is easy to see that
the reflected  radiation is then given by equation  (\ref{eq:lrconv})
with the response function  (\ref{eq:resp}). 

A specific form of the function $\rhoe$ depends on the properties of
the reflecting medium. If, for example, the medium is homogeneous, i.e. 
the single-scattering   albedo  (the  ratio  of   Thomson
cross-section  to the sum of Thomson  and  photoelectric  cross-sections),
$\lambda=\sigmat/(\sigmaph+\sigmat)$, is constant with depth, 
the reflection Green function  $\rhoe$ can be presented as a product of two
Ambarzumian functions (see e.g.  Chandrasekhar 1960; Sobolev 1975)
\be \label{eq:rhophi}
\rhoe(\eta,\zeta)= \frac{\lambda}{4}
\frac{\Hlam(\eta) \Hlam(\zeta)}{\eta+\zeta},
\ee
and   is normalised to the reflection albedo:
\be
\int_0^1 \rhoe(\eta,\zeta) 2 \eta \d \eta = \alam(\zeta)= 
1- \Hlam(\zeta)\sqrt{1-\lambda} .
\ee
For an isotropic source above the infinite plane the angle-averaged albedo then
\be \label{eq:avealb}
\alam=\intl_0^1 \alam(\zeta)\d \zeta=1-h_0\sqrt{1-\lambda},
\ee
where $h_0=2(1-\sqrt{1-\lambda})/\lambda$ is the zeroth moment of $\Hlam$.
The function $\Hlam$ satisfies the integral equation
\be
\Hlam(\eta)=1+\frac{\lambda}{2} \eta
\Hlam(\eta) \intl_0^1 \frac{\Hlam(\zeta) \d \zeta}{\eta+\zeta} .
\ee
For $\lambda\ll 1$, 
\be 
\Hlam(\eta)\approx 1+\frac{\lambda}{2}\eta\ln \frac{1+\eta}{\eta}, \quad
\alam\approx \frac{\lambda}{4} . 
\ee 
One should note that the response  function 
depends on the photon energy  through the function $\rhoe$.

\subsection{Isotropic approximation} \label{sec:isorefl}

The response  function in general  depends on the  properties  of the Green
function  for  reflection  which is a function  of the photon  energy and 
possibly radius (see
eq.~\ref{eq:resp}).  However,
 if the physical conditions do not change with radius and 
if one  assumes that the intensity of the  reflected  
radiation is isotropic (we  will  call  this the isotropic approximation), 
one can approximate $\rhoe(r,\eta,\zeta)=\alam$. 
In this case, one can introduce a
response function that is independent of albedo and photon energy:
\be \label{eq:respiso} 
T(t)\equiv \frac{T_E(t)}{\aE}= 
\int r \d r \intl_0^{2\pi}  \delta(t-\Delta t) \d \phi 
\frac{\zeta}{p^2\cos\xi} \frac{\Omega(\mu)}{\Omega(\cosi)}  
\frac{2 \eta}{2\pi} .
\ee
The  advantage  of this  approximation  is that  we  need to
compute only one response function.  In most
applications  considered  in  this  paper  we use  this  approximation.  We
discuss its accuracy in \S~\ref{sec:acciso}.

\subsection{Integration over azimuth and radius} \label{sec:intazi}

In order to integrate expressions (\ref{eq:resp}) and (\ref{eq:respiso}) over the 
azimuth one can use the following identity:
\beq \label{eq:Gr}
G(r,t)&\equiv&\intl_0^{2\pi}  \delta(t-\Delta t) \d \phi \nonumber \\
&=& \frac{2}{r\sini\ |\sin\phi|} \\
&=&
\left\{ \begin{array}{ll} 
2/\sqrt{r^2\sin^2i -[t-p+(z-h)\cosi]^2} , & i\ne0, \\
2\pi \delta(t-p+z-h), & i=0 . 
\end{array}\right. \nonumber
\eeq
For given $t$ and $r$, the azimuth $\phi$ is found from 
equation~(\ref{eq:delay})
(note that $p$ and $z$ are functions of $r$ only)
\be \label{eq:cosphi}
\cos\phi=-\frac{t-p+(z-h)\cosi}{r\sini} . 
\ee
The limits $\rmin, \rmax$ in  equation~(\ref{eq:Lrefl})  are  determined by
the    conditions    that        expression    under   square   root   in
 (\ref{eq:Gr})      is      positive      (i.e.     $|\cos\phi|<1$,
eq.~\ref{eq:cosphi})  and  that  $\Rin\le  \rmin < \rmax  \le  \Rout$.  The
details  on  the  method  of  integration   over  the  radius  in  equation
(\ref{eq:Lrefl}) are given in the Appendix.


\section{Temporal characteristics} \label{sect:temp}
 
As we showed in  \S~\ref{sec:cosp}  and  \S~\ref{sec:thom},  the  reflected
signal depends linearly on the intrinsic  (direct) signal when there is no
spectral  variability  of  the  direct  radiation  or at  sufficiently  low
energies where Compton down-scattering can be ignored.  Timing properties of
the reflected  radiation are then described by a response  function $T_E(t)$.
For most of our  applications,  we assume that albedo does not depend on radius
and one can take $T_E(t)=\aE T(t)$, where now $T(t)$ is energy independent. 
The  total  observed  radiation  flux is  composed  of the  direct
radiation from the X-ray source plus the radiation  reflected from the disc
(see eq.~\ref{eq:ltot}):
\be \label{eq:total}
\IE(t)=\IDE(t)+\aE \intl_{-\infty}^{t} T(t-t') \IDE(t') \d t' .
\ee
The cross-correlation function (CCF) of the total signal, 
$C_{12}(t)=\int  L_1(t') L_2(t'+t)\d t'$, 
can be easily found substituting equation~(\ref{eq:total}):
\beq
C_{12}(t)&=&\CD_{12}(t) + a_1 a_2 \intl_{-\infty}^{\infty}
\CD_{12}(t') \RT(t-t') \d t' \nonumber \\
&+& \intl_{-\infty}^{\infty}  \CD_{12}(t') \left[a_2 T(t-t')+a_1 T(t'-t)\right] \d t',
\eeq
where $\CD_{12}(t)$ is the CCF of the direct signal and 
$\RT(t)=\int T(t') T(t'+t) \d t'$ is
the auto-correlation of the response function.
The auto-correlation function (ACF) can be obtained 
from the above expression:
\beq
\RE(t)&=&\RDE(t)
+ \aE^2 \intl_{-\infty}^{\infty}  \RDE(t') \RT(t-t') \d t'
\nonumber \\
&+&  \aE \intl_{-\infty}^{\infty} \RDE(t') [T(t'+t)+T(t'-t)] \d t' ,
\eeq
where $\RDE(t)$ is the ACF of the direct signal.

The Fourier transform of (\ref{eq:total}) reads
\be\label{eq:iefou}
\hI_E(f)=\hID_E(f) [1+\aE \hT(f)] .
\ee
The cross-spectrum of the light curves at two energies $E_1, E_2$, 
$\hC_{12}(f)\equiv  [\hI_1(f)]^*\hI_2(f)$, can be expressed as a product 
\be \label{eq:crsp}
\hC_{12}(f)
=\hC^{\rm D}_{12}(f) \hC^{\rm R}_{12}(f) = 
|\hC^{\rm D}_{12}(f)|  e^{i\varphi_{\rm D}(f)}
|\hC^{\rm R}_{12}(f)| e^{i\varphi_{\rm R}(f)} ,
\ee
where $\hC^{\rm D}_{12}(f)=[\hI^{\rm  D}_1(f)]^*\hID_2(f)$  is  
the  cross-spectrum  of the direct  radiation  and 
\be 
\hC^{\rm R}_{12}(f)  \equiv
 1+a_1 a_2 |\hT(f)|^2+a_1 \hT^*(f) +a_2 \hT(f)  .
\ee
The phase  lags  of  the  direct  radiation $\varphi_{\rm  D}(f)$  
and that  added  by  reflection $\varphi_{\rm  R}(f)$ 
are defined as the phases of $\hC^{\rm D}_{12}(f)$ and 
$\hC^{\rm R}_{12}(f)$, respectively.  
The ``reflection'' phase lag, 
\be \label{eq:tanphi}
\tan \varphi_{\rm R}(f) = \frac{(a_2-a_1)\Im \hT(f)}
{1+(a_1+a_2)\Re \hT(f) + a_1 a_2 |\hT(f)|^2} ,
\ee
is positive  (hard lags) when  $a_2>a_1$.  Here $\Re$ and $\Im$ denote real
and  imaginary  part of a complex  variable.  For  small  $a_1$  and  small
$a_2\int T(t)\d t$, the phase lag is approximately:
\be \label{eq:phia2}
\varphi_{\rm R}(f) \approx (a_2-a_1) \Im \hT(f).
\ee
The power density spectra (PDS) are related in the following way:
\be\label{eq:iepds}
|\hI_E(f)|^2=|\hID_E(f)|^2
\left[1+\aE^2|\hT(f)|^2+2\aE \Re \hT(f) \right].
\ee
In many recent publications, the PDS is often 
normalised to the relative rms of the signal (Miyamoto et al. 1991; 
Nowak et al. 1999a). We denote such a PDS $P(f)$.  
In this normalisation, we get 
\be 
P(f)=P^{\rm D}(f)\frac{1+\aE^2|\hT(f)|^2+2\aE \Re \hT(f)}
{\left[1+\aE \int T(t)\d t\right]^2}. 
\ee

\begin{figure*}
\centerline{\epsfig{file=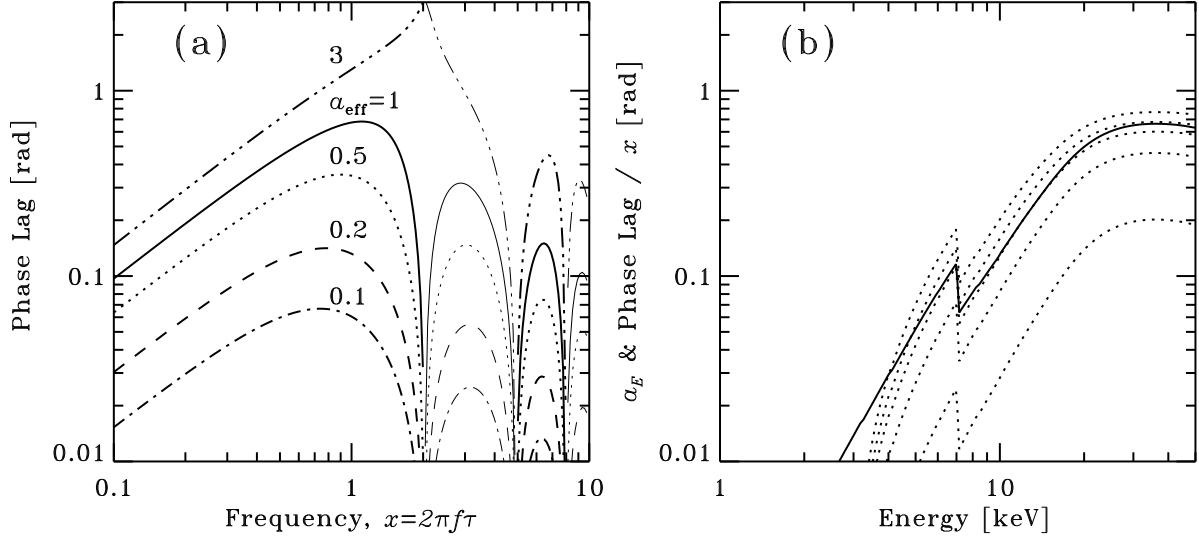,height=7cm}}
\caption{Phase lags  for the exponential response function (\ref{eq:exptr}).
(a)  $\varphi^{\rm  R}_E(f)$ frequency dependence for different $\aeff$  
relative to 3 keV (where $a_1=0.014$). Thin curves give the 
negative phase lags.
(b) Solid curve gives the angle-averaged albedo $\aE$ for Compton  
reflection from neutral medium (omitting Fe K$\alpha$ line)  
corresponding  to the intrinsic photon power-law  index  $\Gamma=2$ 
computed for solar abundances using  {\tt pexrav} model  
(Magdziarz \& Zdziarski 1995) from {\sc   xspec}.  
Dotted curves give the energy  dependence of the phase lag
$\varphi^{\rm  R}_E(f)$ (relative to 3 keV)
divided by frequency  $x\equiv2\pi f\tau$ for $R=1$
at frequencies $x=0.1,0.5,0.7,1,1.5$ (from top to bottom).  At $x<0.1$, the
curves coincide with the $x=0.1$ case.
}
\label{fig:simple}
\end{figure*}

\section{Results} \label{sect:res}

\subsection{Simple response functions}

For  illustration,  let us specify a simple  exponential
response function:
\be \label{eq:exptr} 
T(t)=
\left\{ \begin{array}{ll}
 R\exp[-(t-\tau)/\tau]/\tau, & t\ge\tau,\\
 0, & t<\tau. 
\end{array}  \right.
\ee
The corresponding  Fourier  transform is  $\hT(f)=R\exp(ix)/(1-ix)$,  where
$x=2\pi f \tau$.  As a starting point let us take $E_1=3$ keV where for the
neutral  reflector of solar abundances the albedo $a_1\sim 0.014$.  This is
the smallest  energy  available for timing  analysis, for example, at \xte.
For $\aE\gg a_1$, the phase lag at energy $E$ relative to $E_1$ is then
\be \label{eq:phiaeff}
\tan \varphi^{\rm R}_E(f) \approx   \frac{\aeff (x \cos x+ \sin x)}
{1+x^2+\aeff (\cos x- x \sin x)},   
\ee
where  $\aeff=R\aE$ is the effective  albedo.  The phase lag grows linearly
with frequency as  $\varphi^{\rm  R}_E(f)\approx  2\aeff  x/(1+\aeff)$  for
$x\ll 1$, and for  $\aeff\simlt  3$ reaches the  maximum  $\tan\varphi^{\rm
R}_{\max,1}\approx   0.7   \aeff/(1-0.15\aeff)$   at   $x\approx   1$  (see
Fig.~\ref{fig:simple}a).   The   peak   shift   slightly   towards   higher
frequencies at larger $\aeff$.  When $\aeff\simgt 3$ (large effective
albedo is possible in strongly anisotropic sources), phase lag goes through
$\pi$ at $x\sim 2$.  For  $a_1\ll\aeff\ll  1$ and $x\ll 1$, the phase  lags
are  proportional to the albedo $\aE$ (see  eq.~\ref{eq:phia2}),  and their
energy  dependence  is exactly the same as that  $\aE$.  At small  energies
close to $E_1$, lags go to zero.  At larger energies (and larger  albedos),
the phase lag rises  slower than $\aE$  (Fig.~\ref{fig:simple}b).  At large
frequencies, when $\cos x\sim x\sin x$ (i.e.  $x\sim 1$), the dependence on
$\aE$  in  the   denominator  of   eq.~(\ref{eq:phiaeff})   disappears  and
$\varphi^{\rm  R}_E$ depends  linearly on $a_E$.  Including the fluorescent
iron line  would  produce a bump in the phase  lag  energy  spectrum  which
amplitude depends on the energy resolution of the X-ray instrument.

If the response function is a sum of two exponentials described by 
$\tau_1, R_1$ and $\tau_2, R_2$,  respectively  (let $\tau_1 \ll  \tau_2$), the phase
lag has two  prominent  maxima at $f_1\sim  1/(2\pi  \tau_1)$  and 
$f_2\sim 1/(2\pi \tau_2)$.  The value of the high frequency  maximum at $f_1$ is the
same as for the case of a single  reflector,  while the new, low  frequency
maximum  is  
$\tan \varphi^{\rm   R}_{\max,2}  \approx  0.7  a_{\rm  eff,2}/ [1+a_{\rm eff,1}-0.15a_{\rm eff,2}]$.

Let us now consider the effect of  reflection on the shape of the auto- and
cross-correlation  functions.  The  effect is easier to  understand  if one
considers  an  even  simpler  $\delta$-function   response   
$T(t) = R \delta(t-\tau)$.  Then
\be 
\RE(t)= (1+a_E^2R^2)\RDE(t)+\aE R [\RDE(t-\tau) + \RDE(t+\tau) ] .
\ee
For  monotonic,  strongly  peaked at zero lag ACF (such as that of Cyg X-1,
see Maccarone, Coppi \& Poutanen 2000) and $\aE\ll 1$, the largest relative
increase  in the  total  ACF  achieved  at lag  $\sim  \tau$ is  
$\sim 1+\aE R/\RDE(\tau)$  
(for the   normalised  to unity at zero lag ACF). This means
that the  changes  are  more  significant  when  the  characteristic  decay
time-scale  of the ACF is smaller than or comparable to the  characteristic
delay due to reflection.  Similar behaviour is expected for the exponential
response function (\ref{eq:exptr}).

The CCF for $a_1=0$ and $\delta$-function response  takes the form:
\be 
C_{12}(t)=\CD_{12}(t)+a_2 R\CD_{12}(t-\tau) .
\ee

Since the CCFs of most GBHs and AGNs are  strongly  peaked at zero lag (see
e.g.  Nolan et al.  1981;  Papadakis \& Lawrence 1995; Smith \& Liang 1999;
Lee et al.  2000;  Maccarone et al.  2000), the total  observed  CCF should
become more asymmetric for larger albedos (i.e.  at larger  energies).  The
largest  relative  deviation  from the intrinsic  CCF should be observed at
lags    $t \sim    \tau$. In   the    case    of   the    intrinsic    CCF
$\CD_{12}(t) =\exp[-(t/t_0)^{\nu}]$,  the  relative  change  at  large lags,
$|t| \gg \tau$, is the following:
\be 
\frac{C_{12}(t)}{\CD_{12}(t)}=1+a_2 R
\exp\left[ {\rm sgn} (t)\nu \frac{\tau|t|^{\nu-1}}{t_0^{\nu}}  \right] .
\ee

\subsection{Flat disc response} \label{sec:flat}

\subsubsection{Infinite slab}
\label{sec:infslab}

The response function for a flat disc and an isotropic source can be derived
analytically from equation (\ref{eq:respiso}) 
in the isotropic approximation (i.e.  $\rho=\alam$).  
In this case, $z=0$, $\cos\xi=1$, $\eta=\cosi$, $\zeta=h/p$, and
\be \label{eq:flatint}
T_{\rm slab}(t)=\frac{2\cosi}{2\pi} \int \frac{r\ \d r}{p^2} \frac{h}{p} G(r,t),
\ee
where $p^2=r^2+h^2$ and  $G(r,t)$ is given by equation~(\ref{eq:Gr}). 
The variable $p$ should satisfy the condition   $p_-<p<p_+$, where
\be \label{eq:ppm}
p_\pm=q\pm s, \ q=\frac{t-h\cosi}{\cos^2 i},
\  s=\frac{\sini \sqrt{t(t-2h\cosi)}}{\cos^2 i}.
\ee
Substituting $p=q+s\cos\theta$ into equation~(\ref{eq:flatint}) we get:
\beq \label{eq:tflat0}
T_{\rm slab}(t)&=&\frac{2\cosi}{2\pi} \intl_{p_-}^{p_+} \frac{h\d p}{p^2}
\frac{2}{\cosi \sqrt{(p-p_-)(p_+-p)}}   \nonumber \\
&=& \frac{2h}{\pi} \intl_0^\pi \frac{\d\theta}{(q+s\cos\theta)^2} =
\frac{2h}{\pi} \frac{\pi q}{(q^2-s^2)^{3/2}}    \\
&=& 2\cosi \frac{h(t-h\cosi)}{(t^2+h^2-2th\cosi)^{3/2}}, \quad 
t\geq 2h\cosi . \nonumber
\eeq
The response function is normalised in the following way:
\be \label{eq:flnorm}
\intl_{2h\cosi}^{\infty} \ T_{\rm slab}(t) \d t =2\cosi.
\ee

\begin{figure*}
\centerline{\epsfig{file=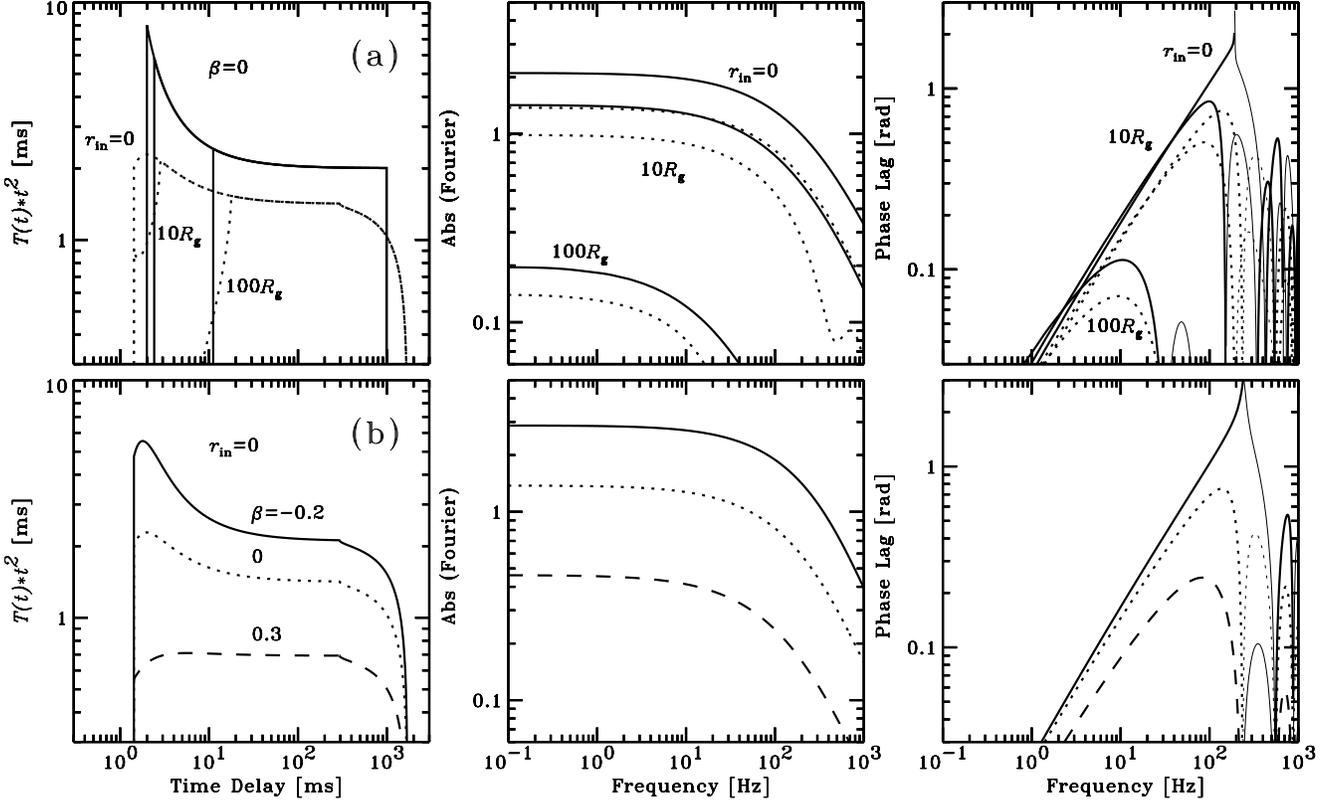,width=17.5cm}}
\caption{(a) Response  functions $T(t)$ (in the isotropic  approximation),  
their Fourier  amplitudes, and the phase lags for
the flat disc with  different  inner  radius  $\Rin$ and $\Rout=10^4\Rg$. 
Since  asymptotically at large $t$  in flat discs $T(t)\sim  t^{-2}$,
the responses are  multiplied by $t^2$.   
The  phase  lags are computed  assuming  albedos  $a_1=0, a_2=1$.  
Solid and dotted  curves correspond to  $\cosi=1$  and  $0.7$,  respectively.   
Thin curves give negative phase lags.  
(b) Dependence on the bulk velocity $\beta=v/c$. 
Solid, dotted, and dashed curves correspond to $\beta=-0.2, 0, 0.3$, 
respectively. Parameters are $\Rin=0$, $\Rout=10^4\Rg$, $\cosi=0.7$.
The negative $\beta$ means the bulk velocity directed towards the disc.  
$\Rg/c=10^{-4}$ s in all simulations.}
\label{fig:flat}
\end{figure*}

\subsubsection{Disc with a central hole} 

In hot flow models for the X-ray  production in accreting  black holes, the
X-ray source is situated  inside a cold  accretion  disc  truncated at some
radius (see e.g.  Esin et al.  1998; Poutanen 1998).  The response then can
be approximately  evaluated assuming a point source at the axis of the disc
with $\Rin\gg1$ (see also Gilfanov,  Churazov \& Revnivtsev  2000b). 
A more general expression for the response function,
for a non-zero inner radius and a finite outer radius $\Rout$, 
can be written similarly to equation  (\ref{eq:tflat0}):
\beq \label{eq:tflatmin}
T(t)&=& \frac{2h}{\pi} \intl_{\pmin}^{\pmax}
\frac{\d p}{p^2\sqrt{(p-p_-)(p_+-p)}}   \nonumber \\
&=& \frac{2h}{\pi}  \intl_{\thmax}^{\thmin} 
\frac{\d\theta}{(q+s\cos\theta)^2},  \quad \tmin<t<\tmax ,
\eeq
where  
\beq
\tmin &=& \max(2h\cosi,\sqrt{\Rin^2+h^2}+h\cosi-\Rin\sini) , \nonumber \\
\tmax &=& \sqrt{\Rout^2+h^2}+h\cosi+\Rout\sini ,
 \eeq
and the limits     
\beq
\pmin &=& \max( \sqrt{\Rin^2+h^2}, p_-), \nonumber \\ 
\pmax &=& \min( \sqrt{\Rout^2+h^2}, p_+), \\   
\theta_{\rm m,M} &=& \arccos[(p_{\rm m,M}-q)/s] . \nonumber 
\eeq
For  $\sqrt{\Rin^2+h^2}+h\cosi+\Rin\sini \le t \le\sqrt{\Rout^2+h^2} +h\cosi-\Rout\sini$,   
the   response   function   is  given  by   equation (\ref{eq:tflat0}). 
Fig.~\ref{fig:flat}a shows the response function and other timing  characteristics
for different inner radii $\Rin$.  Since in flat discs   $T(t)\sim  t^{-2}$ at large $t$ 
(see eq.~[\ref{eq:tflat0}]), the responses are  multiplied by $t^2$.  
Due to the absence of the  reflector in the direct  vicinity of the flare, 
reflection acts as a low  pass  filter  removing  high  frequency   signal  
from  the  reflected radiation.  This  causes  the  maximum  of the phase  lag 
to shift to lower frequencies.  For large  $\Rin$, the covering  factor of the  
reflector  is small and therefore the amplitude of the phase lag decreases.

\begin{figure*}
\centerline{\epsfig{file=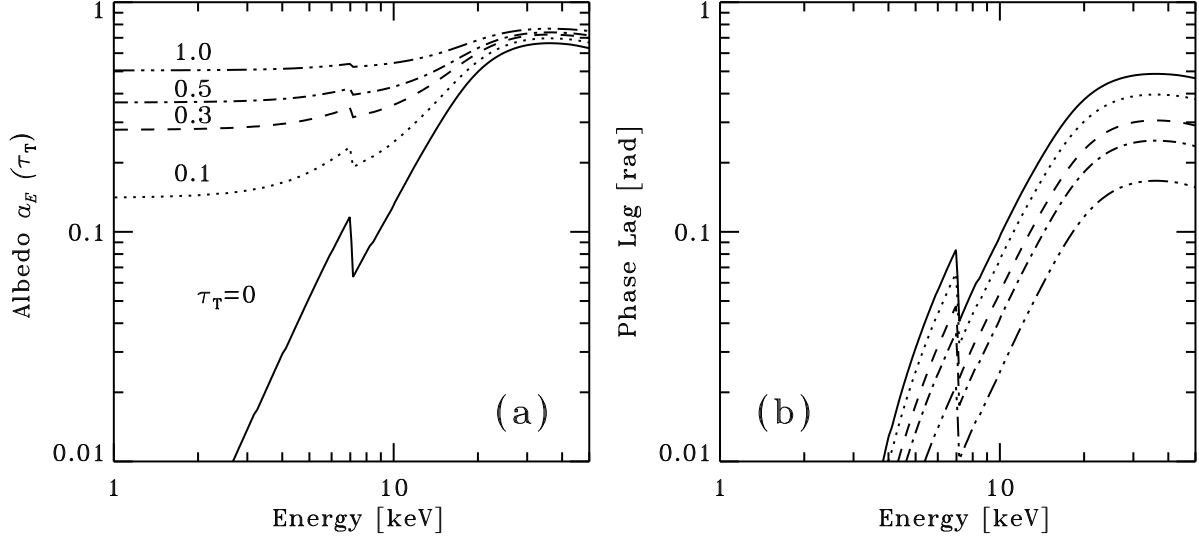,height=7cm}}
\caption{
(a) Energy  dependence  of the  reflection  albedo  for a slab with  purely
scattering  skin of Thomson  thickness  $\tT$ atop of a neutral slab.   
(b) Energy  dependence  of the phase lags.  An  isotropic  X-ray source is
elevated at $h=10$ above an infinite  ``ionised''  slab.  The phase lags are
computed  relative to $E_1=3$ keV at Fourier  frequency  $f=1/(10 h)$ (i.e.
100 Hz for $h=10\Rg= 3\ 10^{7}$ cm).
}
\label{fig:ion}
\end{figure*}

\begin{figure*}
\centerline{\epsfig{file=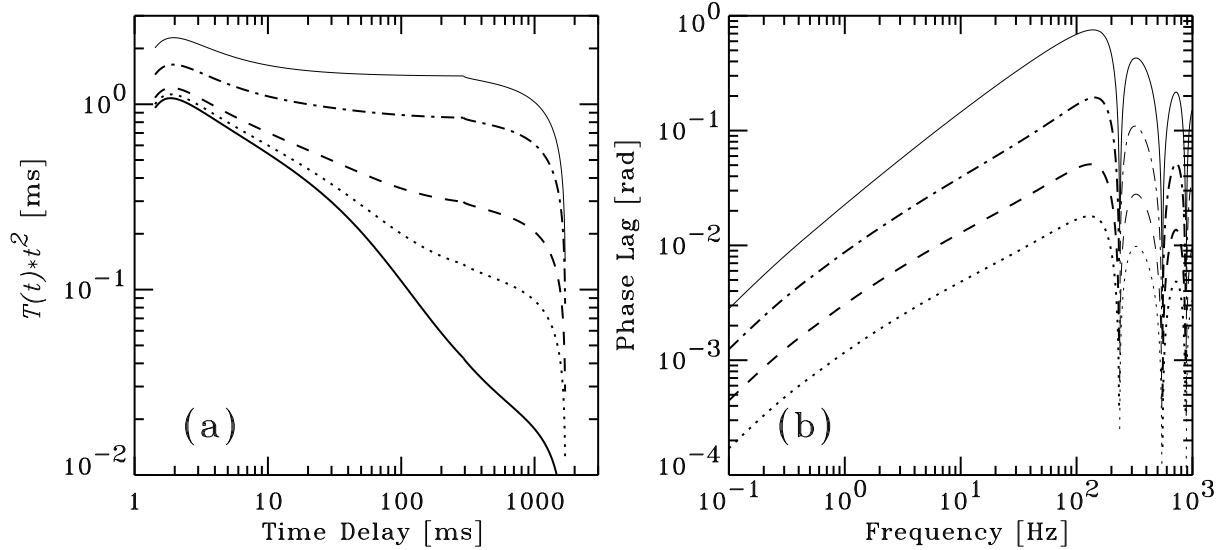,height=7cm}}
\caption{
(a) Response functions $T_E(t)$ for a flat slab with ionised skin 
which optical depth decreases with radius as $\tau_T(r)=\tau_0 (1+ r/10h)^{-3/2}$, 
where $\tau_0=1$, and $h=10$ are assumed (here $\cosi=0.7$ and $\Rg=3\ 10^6$ cm).
Thick curves from the bottom to the top correspond to 3, 6, 12, 24 keV.
Thin solid curve represents the response function with unity albedo
over the whole surface.   
(b) Corresponding phase lags relative to 3 keV. Thin solid curve 
represents the case of constant albedos $a_2=1, a_1=0$. 
}
\label{fig:ion2}
\end{figure*}

\subsubsection{Anisotropic sources}

Let us consider  anisotropic sources of radiation.  When the X-ray emitting
plasma has some bulk velocity  $\beta\equiv  v/c$ away from or towards the disc,
the angular distribution of the radiation is
\be \label{eq:bulk}
\Omega(\mu)=\frac{1}{\gamma^4(1-\beta\mu)^3} , \quad \gamma=1/\sqrt{1-\beta^2},
\ee 
if  the  emission  is  isotropic   in  the  plasma  rest  frame  (see  e.g.
Rybicki \& Lightman 1979).  
This  formula  is  valid  for  both  the   bolometric
luminosity  as  well  as for  the  monochromatic  luminosity  for a  photon
spectral index $\Gamma=2$. 
Similarly to derivations in \S~\ref{sec:infslab}, 
we then can analytically compute the infinite slab response function 
substituting $\Omega(-h/p)$ to equation (\ref{eq:respiso}) :
\beq \label{eq:betaresp}
&T_{\beta}(t) =& \frac{2h}{\pi} (1-\beta\cosi)^3 
\intl_0^\pi \frac{\d\theta}{p^2 (1+\beta h/p)^3}   \nonumber \\
&=&  \frac{2h}{\pi} (1-\beta\cosi)^3  \sum_{k=0}^{\infty}  
 \frac{(k+1)(k+2)}{2} 
(-\beta h)^k  \nonumber \\ 
&\times & \intl_0^\pi \frac{\d\theta}{(q+s\cos\theta)^{k+2}} 
\nonumber \\
&=& h(1-\beta\cosi)^3\sum_{k=0}^{\infty} (k+1)(k+2) (-\beta h)^k \\
&\times &\frac{P_{k+1}\left( q/\sqrt{q^2-s^2}\right) } 
{(q^2-s^2)^{1+k/2}} , \nonumber
\eeq
where $P_k$ are the Legendre  polynomials and we used 
relation 3.661 from Gradshteyn \& Ryzhik (1980) to obtain the 
last expression. The responses together with the
related phase lags are presented in Fig.~\ref{fig:flat}b for a few $\beta$.
With increasing  $\beta$, the response amplitude  decreases since radiation
is becoming more beamed away from the disc towards the  observer.  At small
lags the effect is even more dramatic.  This causes also a significant change
in  the amplitude of the phase lags at high frequencies.

\subsubsection{Ionised discs}

The  intense  incident  X-ray flux can affect the  ionisation  state of the
reflecting   material  changing  its  albedo.  The  surface  layer  of  the
accretion disc is heated up to the Compton  temperature of radiation and is
almost completely  ionised (e.g.  Nayakshin,  Kazanas \& Kallman 2000).  We
can consider  such  ``skin'' as a purely  scattering  medium.  A transition
layer from the hot layer to almost neutral matter is very thin.  Therefore,
a  two-phase  model,  with a hot layer  atop of  neutral  material,  can be
adopted.  We neglect smearing of the reflection features by thermal motions
in the skin and consider pure Thomson scattering.

One should point out here that in the isotropic approximation, the response
functions $T(t)$ given by equation (\ref{eq:respiso})
are not affected by  ionisation (while the reflected luminosity 
is affected).  However,  the phase lags at given
energies  will  change  since  they  depend on the  reflection  albedo.  If
reflection from the underlying  neutral material follows Lambert's law (i.e
isotropic  intensity),  the total  reflection  albedo can be  expressed  as
follows (Sobolev 1975):
\be \label{eq:albang}
a(\zeta,\tT)=1-2(1-a)
\frac{1+\frac{3}{2} \zeta + \left(1-\frac{3}{2}\zeta\right) 
\exp(-\tT/\zeta)}{4+3\tT(1-a)},
\ee
where $\tT$ is Thomson optical thickness  of the  ionised  layer, and 
$a$ is the  albedo for a neutral slab.  The angle averaged albedo is then
\beq \label{eq:atae}
a(\tT)&=&1-(1-a)
\frac{7-3\exp(-\tT)+(4+3\tT)E_2(\tT)}{2[4+3\tT(1-a)]} 
\nonumber \\ &\approx& 
1-(1-a)\frac{7+\exp(-10\tT)}{2[4+3\tT(1-a)]} ,
\eeq
where $E_2$ is the exponential integral of the 2nd order.
A relative error of the  approximation in the nominator is smaller than 1\%
for any $\tT$.  The energy  dependence of the albedo for different $\tT$ is
shown in  Fig.~\ref{fig:ion}a.  We see that at lower energies the albedo is
high,  $a_{\min}  (\tT) \approx  3\tT/(4+3\tT)$,  compared with the neutral
material  albedo, and therefore one cannot assume $a_1\ll 1$ when computing
the phase lags.  The  important  quantity  here is, however, not the albedo
itself, but the  difference  between  albedos at  different  energies  (see
eqs.~\ref{eq:tanphi}, \ref{eq:phia2}):
\beq\label{eq:albdif}
a_2(\tT)-a_1(\tT)\!\!&\approx&\!\! 
\frac{(a_2-a_1)\{1-[1-\exp(-10\tT)]/8\} }
{[1+3\tT(1-a_1)/4][1+3\tT(1-a_2)/4]} 
\nonumber \\
&\approx&\!\! \frac{a_2-a_1}{(1+3\tT/4)^2}.
\eeq
The first approximate formulae is accurate within 1\%.  The accuracy of the
second one is better than 13\% for $\tT<0.5$,  and is better than 20\% upto
albedo $a_2\sim 0.7$ (i.e.  $E\sim 20$ keV) for $\tT\sim 1$.  The resulting
phase lags  (relative  to 3 keV)  at  $f=1/(10h)$ as a
function  of photon  energy  are shown in   Fig.~\ref{fig:ion}b.  They have
very similar behaviour at other frequencies (see Fig.~\ref{fig:simple}) and
follow the $a_2(\tT)-a_1(\tT)$ dependence (see eq.~\ref{eq:tanphi}).

In real physical situations, the optical depth of the ionised skin could 
be a function of the radius. For a source 
of ionising radiation at the disc axis at height $h$, 
we can approximate the radial distribution  as
$\tT(r)=\tau_0 (1+r/10h)^{-3/2}$ (see e.g. Nayakshin 2000). 
The resulting response functions $T_E(t)$ computed from equation (\ref{eq:resp}) 
in isotropic approximation are shown in Fig.~\ref{fig:ion2}a. 
Close to the center (small delays), the responses for different energies 
are close to each other, due to the fact that the effective albedo is large 
for all energies (see eq.~[\ref{eq:atae}]). At large delays, a strong
energy dependence of the effective albedo results in large differences in the 
responses. The response is strongly suppressed at small energies, 
while at higher energies the shape of the response function is similar to 
the constant albedo case. 
The resulting phase lags  (see Fig.~\ref{fig:ion2}b) 
are then smaller (comparing to the constant albedo case) at higher frequencies  
since the difference of the albedos becomes smaller (see eqs.~[\ref{eq:tanphi},
\ref{eq:phia2}, \ref{eq:albdif}]). Phase lags at low frequencies are 
hardly affected by ionisation.

\subsection{Flared disc response}

\begin{figure*}
\centerline{\epsfig{file=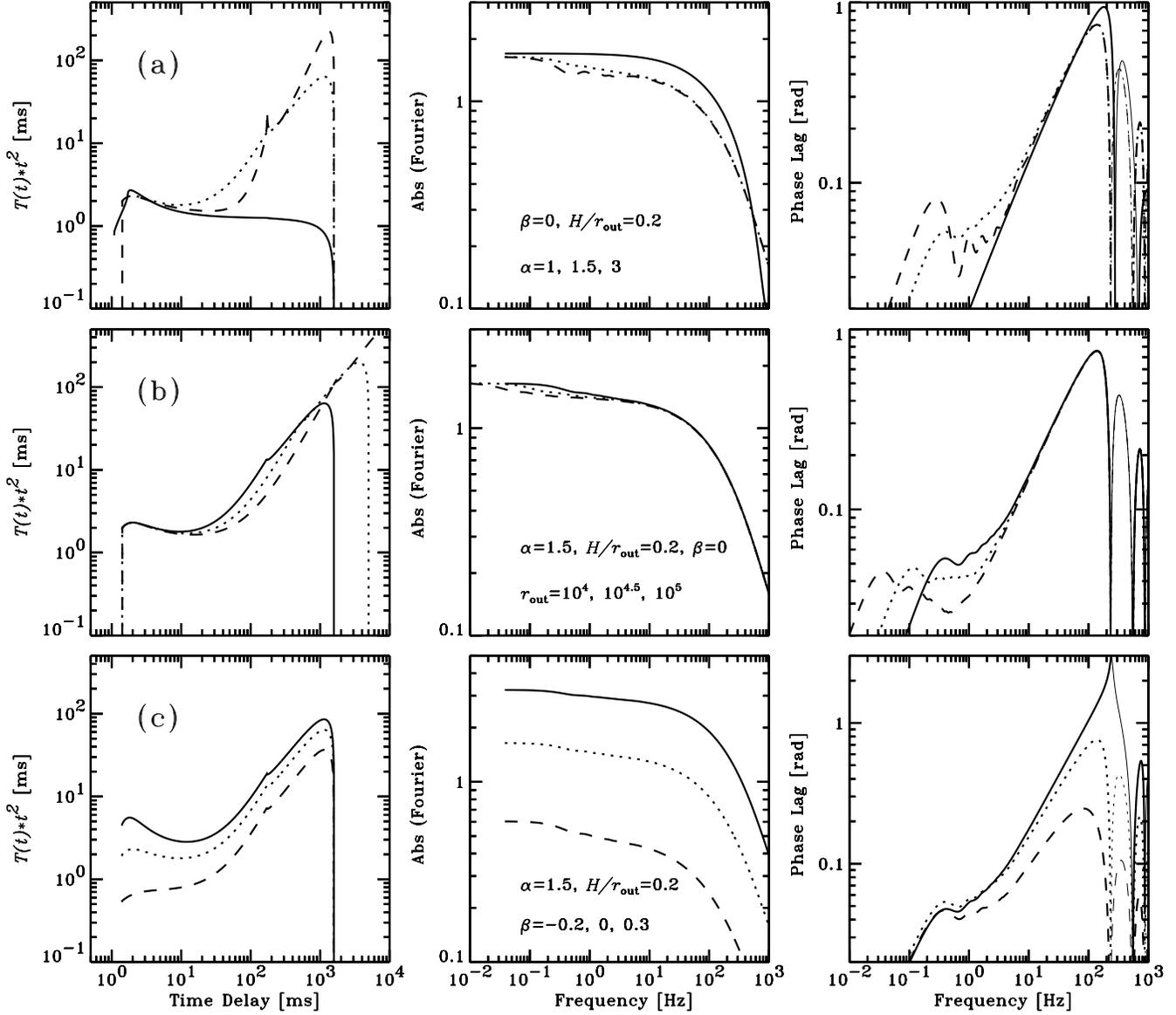,width=17.5cm}}
\caption{Response  functions,  their Fourier  amplitudes, and the phase lags 
for flared discs in the isotropic approximation.
The  phase  lags are computed  assuming  albedos  $a_1=0, a_2=1$.  
Thin curves give negative phase lags. 
(a) Dependence on $\alpha$. Solid, dotted, and dashed curves 
correspond to $\alpha=1, 1.5, 3$, respectively.   
(b) Dependence on the outer disc radius $\Rout$. Solid, dotted, and 
dashed curves correspond to $\Rout=10^4, 10^{4.5}, 10^5\Rg$, respectively.
(c) Dependence on the bulk velocity $\beta=v/c$. Solid, dotted, and 
dashed curves correspond to $\beta=-0.2, 0, 0.3$, respectively.  
Everywhere $\cosi=0.7$, and in (a), (c)  $\Rout=10^4\Rg$.
}
\label{fig:flared}
\end{figure*}

An important  parameter influencing  the response of a flared disc is the
$\alpha$-parameter (see eq.~\ref{eq:z}) which describes the disc shape. 
In the general case, it is difficult to get useful analytical formulae for the 
disc response, therefore we compute the responses numerically (see eq. 
[\ref{eq:respiso}] and Appendix). 
We first  consider  an  isotropic  source  above the  disc  and  vary  the
$\alpha$-parameter.  The results are shown in  Fig.~\ref{fig:flared}a.  For
$\alpha=1$  the  response  does not differ much from that of the flat disc.
For any $\alpha>1$, there appears an increased response from the outer part
of the  disc  and  the  response  function  flattens  out  at  large  lags.
Comparing  with the flat discs (and  flared  discs  with  $\alpha=1$),  the
additional phase lags corresponding to the light crossing time to the outer
part  of the  disc  $\tau\sim\Rout/c$  appears  at low  frequencies  $f\sim
1/(2\pi\tau)$.  The larger the curvature (i.e.  larger  $\alpha$), the more
prominent the bump in the  Fourier-frequency--dependent  phase lag spectrum
becomes.

Changes  of  the  accretion  disc  size  affect  mostly  the  lags  at  low
frequencies   corresponding   to  the   response   from  the   outer   edge
(Fig.~\ref{fig:flared}b).  For larger $\Rout$, the secondary phase lag peak
shifts towards lower frequencies with a constant amplitude.

Variations  in the plasma bulk  velocity and  corresponding  changes in the
angular  distribution of the intrinsic radiation affect markedly the flared
disc  response  functions  and  the  phase  lags.  They  are  presented  in
Fig.~\ref{fig:flared}c for a few $\beta$.  The most prominent effect is the
dramatic  change in the amplitude of the response at small lags, i.e.  from
vicinity  of the X-ray  source.  The  response  from the outer disc edge is
affected less since for  $\eta\sim 0$ and mildly  relativistic  velocities,
beaming is not so  important.  The  resulting  phase lags are affected in a
similar way:  the amplitude of the phase lags at high  frequencies  changes
dramatically,  while the lags at low frequencies  related to the outer disc
edge remain basically the same.

\subsection{Approximations}

\subsubsection{Accuracy of isotropic approximation}
\label{sec:acciso}

\begin{figure*}
\centerline{\epsfig{file=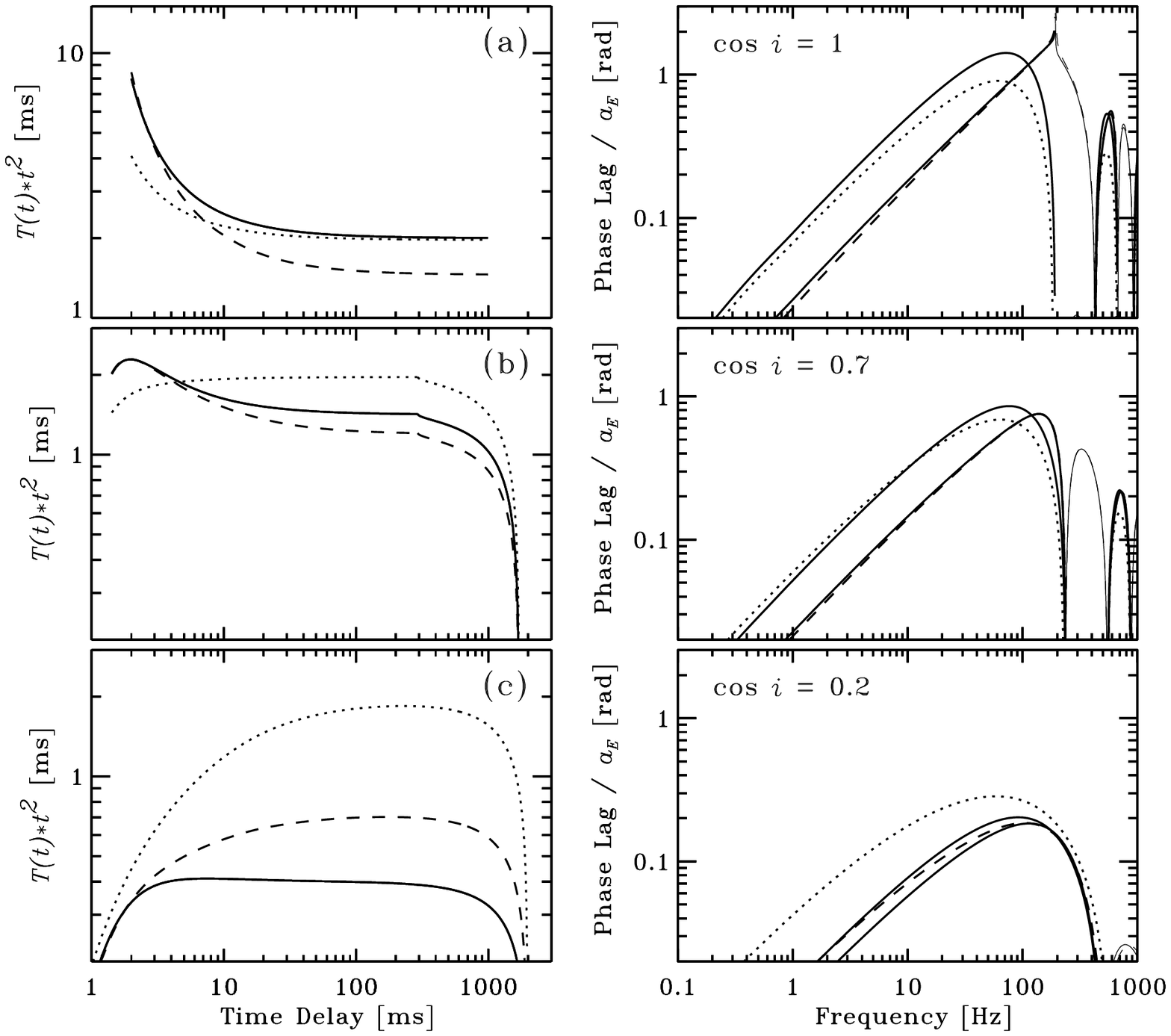,height=12cm}}
\caption{
Response  functions  and  phase  lags  for the  flat  disc  of  radius
$\Rout=10^3$  with an isotropic  source at  elevation  $h=10$  computed for
three  inclinations  $\cosi=1,  0.7, 0.2$ (from top to  bottom).   
Dotted  and  dashed  curves represent the  response  function  $T_E(t)$  given by
equation  (\ref{eq:resp})  computed  using  the  exact  angle-dependent
expression for $\rhoe$ from equation  (\ref{eq:rhophi})  for  $\lambda=0.1$
and 1,  respectively, and divided by the angle-averaged albedo for these 
$\lambda$ (i.e. by $\alam=0.026$ and 1, see eq.~[\ref{eq:avealb}]). 
The  corresponding  phase  lags  (computed  assuming
$a_1=0$) are also divided by $\aE$.  Thin curves give negative
lags.  Solid curves  correspond to the response  function (\ref{eq:respiso}) 
computed assuming isotropic   intensity  of  the  reflected   radiation,  i.e.
$\rhoe=\alam$.  The phase lags are computed taking the same  angle-averaged
albedos, $\alam=0.026$ and 1.  For larger albedo,   the phase
lags peak shifts to higher frequency.
}
\label{fig:lambda}
\end{figure*}

Let us now estimate the accuracy of the isotropic approximation.  Using the
exact expressions for $\rhoe$-function (eq.~\ref{eq:rhophi}) we computed the
response   functions   for  the two cases 
 $\lambda=0.1$   and  1 corresponding to the photoelectric absorption and 
the scattering dominated regimes. The results are  shown  in
Fig.~\ref{fig:lambda}.  The  difference  is coming from  different  angular
dependences of the reflection Green function~(\ref{eq:rhophi}) at different
$\lambda$.

For  small   $\lambda$,   $\Hlam\sim   1$  and   $\rhoe(\cosi,\zeta)/\aE\sim
1/(\cosi+\zeta)$.  At  rather  small   inclinations,   $\cosi=0.7-1$,   the
reflected   intensity  is  reduced   comparing   to  the   isotropic   case
($\rhoe/\aE=1$)  at small lags (photons  coming  underneath the flare) since
$\zeta\sim 1$.  At large delays, $\zeta\sim 0$ and the reflected  intensity
is closer to the isotropic  case.  For small  inclinations,  $\cosi\sim 0$,
the  reflection is  significantly  larger than that given by the  isotropic
approximation  at large delays  (i.e.  small  $\zeta$)  (see lower panel in
Fig.~\ref{fig:lambda}).

For $\lambda\sim 1$, $\Hlam(\eta)  \sim 1+2\eta$ and $\rhoe  (\cosi,\zeta) /
\aE \sim (1+2\cosi)  (1+2\zeta) /4 (\cosi+\zeta)$.  At small  inclinations,
the isotropic  approximation  underestimates  (overestimates)  slightly the
response at small (large) lags.  At large  inclinations  and large  delays,
the response exceeds that in the isotropic approximation.

Thus, the isotropic approximation  overestimates  (underestimates) response
at small (large) lags for most  inclinations  and for $\lambda\ll  1$.  For
$\lambda\sim  1$, the  isotropic  approximation  is very  accurate at small
lags.  At large lags, it overestimates  (underestimates)  slightly response
at small (large) inclinations.

The phase lags  expected for the flat disc and computed  using the accurate
response function as well as in the isotropic  approximation  are presented
on the right  panels  in  Fig.~\ref{fig:lambda}.  For  $\lambda\ll  1$, the
``isotropic''  phase lags are larger for small inclinations and smaller for
large  inclinations than the accurate ones.  For $\lambda\sim 1$, the phase
lags computed in the isotropic  approximation are almost  indistinguishable
from the exact ones  (compare  dashed and solid  curves  peaking  at higher
frequencies).

\subsubsection{Approximate response function} \label{eq:appresp}

\begin{figure*}
\centerline{\epsfig{file=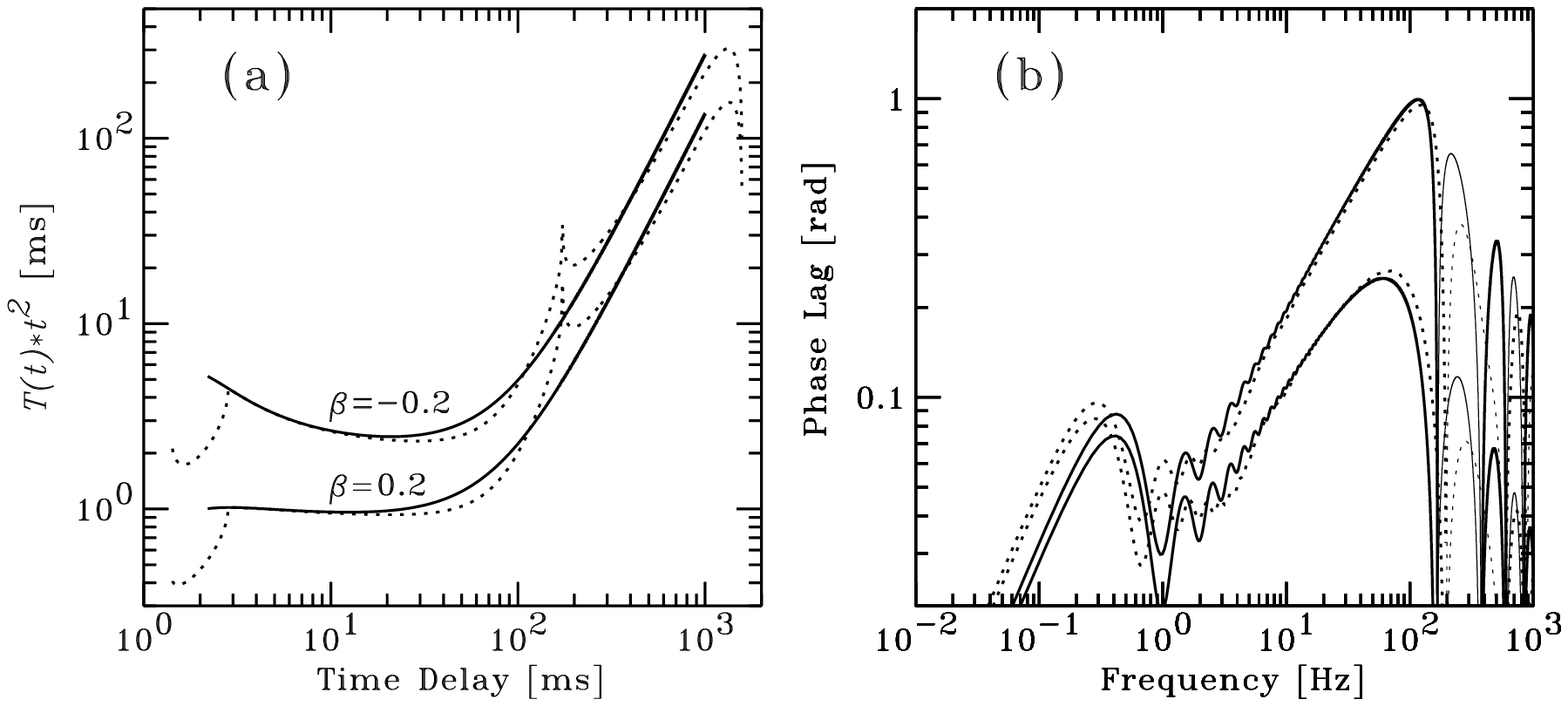,height=7cm}}
\caption{
(a) Flared  disc response  functions for $\alpha=3$,   
$H/\Rout=0.2$, $h=10$, $\cosi=0.7$, $\Rin=10$, and $\Rout=10^4$. 
Dotted curves give the exact responses for $\beta=-0.2$ (upper curve) and
$\beta=0.2$ (lower curve).
Solid curves represent the approximate response function (\ref{eq:tappr2}). 
(b) Corresponding phase lags  (computed for $a_1=0,  a_2=1$). 
}
\label{fig:app}
\end{figure*}

Calculations  of the response  function from a flared disc are not entirely
trivial.  However, for many applications they are not even needed.  One can
approximate the exact results by much simpler response  functions that have
almost  identical  temporal  properties.  The reflection  response from the
vicinity  of the  X-ray  source  can be    approximated  by the flat  disc
response  (\ref{eq:tflat0}),  while the response from the outer edge can be
approximated by a power-law:
\be\label{eq:tappr}
T_{\rm appr}(t)=R_1 T_{\rm slab}(t) + R_2 T_{\rm out}(t)  ,
\ee
where  $T_{\rm out}(t) =c\ t^{-2+1.6(\alpha-1)^{1/3}}$, $\tmin < t < \Rout$,
and the constant  $c$ is found from the normalization condition 
$\int T_{\rm out}(t) \d t=1$.  
We find that  the resulting phase lags are similar to the 
exact ones (for $\Rin\ne0$) if one takes 
\be \label{eq:tmin}
\tmin=
\left\{ \begin{array}{ll} 
\sqrt{\Rin^2+h^2}+h\cosi , & x\ge 1, \\
(1-x)2h\cosi+x(\sqrt{\Rin^2+h^2}+h\cosi), & x<1  ,
\end{array}\right. \nonumber
\ee
where $x=\Rin/h\tan i$. The  normalisation  factors,  i.e.  the  apparent
amplitude  of  reflection  from the disc area close to the source  and that
from the outer edge,  depend on the  inclination,  bulk  velocity  $\beta$,
and solid angle  occupied by the outer disc edge expressed in $H/\Rout$:
\be
R_1= \int_{-1}^{0} \d \mu \frac{\Omega(\mu)}{\Omega(\cosi)}  =
(1-\beta\cosi)^3 \frac{1+\beta/2}{(1+\beta)^2} ,  
\ee
\beq
R_2&=&\frac{H}{\Rout} \int_{0}^{H/\Rout} \d \mu 
\frac{\Omega(\mu)}{\Omega(\cosi)} \nonumber \\ 
&=& 
\frac{H}{\Rout} (1-\beta\cosi)^3 
\frac{1-(\beta/2)H/\Rout}{(1-\beta H/\Rout)^2} .
\eeq
A more accurate approximation can be introduced if one replaces the first
term  in equation  (\ref{eq:tappr})  by the  ``anisotropic''  response
(\ref{eq:betaresp}):
\be\label{eq:tappr2}
T_{\rm appr}(t) = T_{\beta}(t) + R_2 T_{\rm out}(t) . 
\ee
This  approximation  is compared with the accurate  response  functions  for
flared  discs at  Fig.~\ref{fig:app}.  One sees that it
reproduces rather well the response  functions. The approximation 
slightly overestimate (by 30\% -- 50\%) the position of the peak
in the phase lag spectra resulting in a similar error in estimation of 
the disc radius. 

\begin{figure*}
\centerline{\epsfig{file=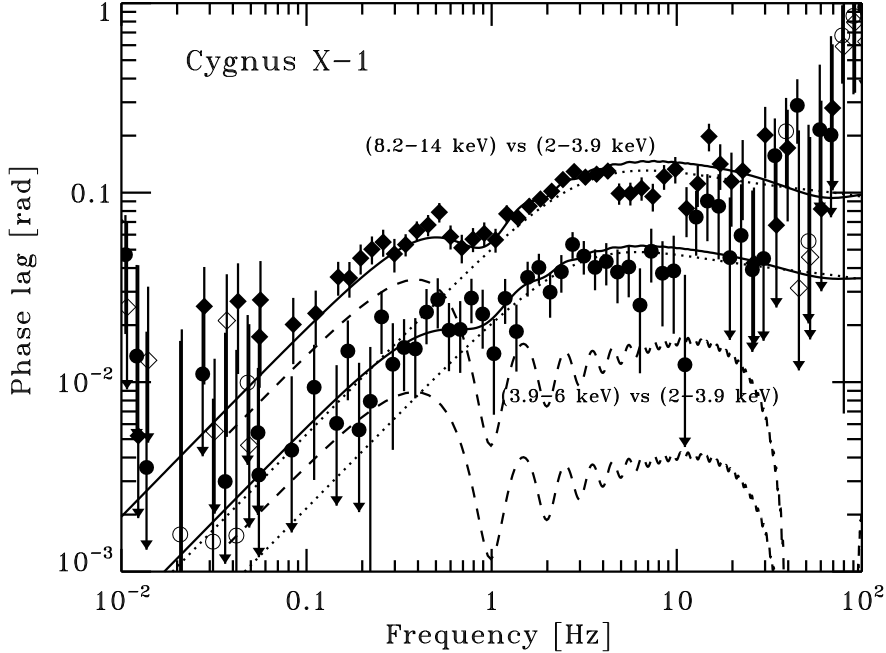,height=9.0cm}}
\caption{
Phase lags between signals in the 8.2-14 keV and the 3.9-6 keV bands vs the
2-3.9 keV band observed with \xte\ in Cyg X-1 on 1996, October 22 (Nowak et
al.  1999a;  Pottschmidt  et al.  2000).  Dotted curves  correspond  to the
intrinsic lags described by a modified shot noise model (Poutanen \& Fabian
1999) with  exponentially  rising  shots of  time-scale  $t_0$  distributed
between 1~ms and 0.05~s according to a power law  $P(t_0)\propto  t_0^{-p}$
with   $p=1.5$.  At  a  given   energy   $E$   the   shot   time-scale   is
$t_E=t_0[1-b\ln(E/E_0)]$,  where  $E_0=3$  keV and  $b=0.22$.  Since  $t_E$
decreases  with  energy,  the lags are  hard.  Dashed  curves  are the lags
produced   by   reflection from a flared disc with the following parameters
$\Rin=10$,  $\Rout=10^3$, $h=10$ (in units $\Rg=3\ 10^6$ cm), 
$H/\Rout=0.3$, $\alpha=3$,  $\cosi=0.7$, and  $\beta=0$.
Solid curves give the sum of the lags from the intrinsic signal and reflection.
}
\label{fig:cygx1lag}
\end{figure*}

\section{Comparison with Observations and Discussion} \label{sect:disc}

The phase lag Fourier spectra of a number of GBHs show breaks and secondary
peaks at low  frequencies  (Nowak et al.  1999a,b;  Grove et al.  1998; see
Fig.~\ref{fig:cygx1lag}).  The  appearance  of a  secondary  peak  could be
related to the variability  properties of the intrinsic  signal.  The peaks
in the phase  lag  spectra  approximately  correspond  to the peaks in the
power density spectra in the $fP(f)$ representation (see e.g.  Nowak et al.
1999a;  Kotov,  Churazov \& Gilfanov  2001)  which can be  produced  by two
independent  processes.  The secondary phase lag peaks can also be a result
of  reflection  from  the  outer  part of the  disc  (Poutanen  2001).  The
position  of  these  peaks  would  then   depend  on  the  disc  size  (see
Fig.~\ref{fig:flared}b).  For a linear response, the energy dependence of
the lags at a given frequency should have features  characteristic  of
reflection:  an Fe line, an edge, and hardening above 10 keV.  The lags 
observed in Cyg X-1 at frequencies  above 0.6 Hz do not
show  these  features  (Kotov  et  al.  2001). In fact, the
reflection features appear to have negative  contribution  (``anti-lags''),
i.e.  the lags are {\it reduced}  at  energies 
where reflection is expected to contribute.  This could be interpreted as a
non-linear,   negative  response.  At  lower   frequencies (below 0.6 Hz),  
the energy dependence of lags is not known. It is possible 
that there the reflection response is still positive and the lag energy 
dependence shows reflection  features.

Let us develop a toy model that is capable of describing the observed phase
lags  (and the  PDS).  We  assume  that the  lags at high  frequencies  are
intrinsic,  i.e.  related to the direct  radiation.  We describe the direct
emission in terms of a simple  modified shot noise model.  The shot profile
is taken to be a rising exponential  $\exp(t/t_E), \ t<0$.  For a power law
radiation  spectrum with  time-varying  spectral index, the shot time-scale
depends logarithmically on energy $t_E=t_0[1-b\ln(E/E_0)]$ (see Poutanen \&
Fabian  1999; Kotov et al.  2001).  Spectral  evolution  can for example be
produced by evolution of magnetic  flares  (Poutanen  \& Fabian  1999).  We
consider  $b$ as a free  parameter  since its  value is  determined  by the
details of spectral  evolution  which are unknown.  The decreasing of $t_E$
with  energy  ensures  that the ACF width  decreases  with $E$ as  observed
(Maccarone  et al.  2000).  The  resulting  CCFs peak at zero lag since the
shots at different energies are correlated and peak at the same time.  For a
given  $t_0$,  the   cross-spectrum  of  the  signal  at  two  energies  is
$\hC(f,t_0)\propto t_1 t_2/[(1-i2\pi f t_1) (1+i2\pi f t_2)]$.  If $t_0$ is
distributed  according  to  $P(t_0)$,  the  total  cross-spectrum   becomes
$\hC(f)=\int \hC(f,t_0) P(t_0) \d t_0$.  We assume a power-law distribution
$P(t_0)  \propto  t_0^{-p}$  between  $t_{\min}$ and  $t_{\max}$.  The lags
observed  from Cyg X-1 on 1996,  October  22 (Nowak et al.  1999a)  at high
frequencies   can  be   described   by  this   model   (dashed   curves  in
Fig.~\ref{fig:cygx1lag}) with $t_{\max}=0.05$~s, $p=1.5$, $b=0.22$, and any
sufficiently  small $t_{\min}$.  At $f<1/(2\pi  t_{\max})$,  phase lags are
proportional to $f$.  With these  parameters, the PDS of Cyg X-1 above 1 Hz
is also well  reproduces  ($p=1.5$ gives the slope of the PDS also equal to
1.5, see Poutanen \& Fabian 1999).  The PDS shape at lower  frequencies can
be reproduced for example by the flare avalanches.

Let us now consider the effect of  reflection  by applying an approximate
response  function for a flared disc given by equation~(\ref{eq:tappr2}).  
A rather good fit to the data can be achieved for the disc inner and 
outer radii $30\Rg$ and $1000\Rg$, respectively, and the flaring parameter 
$H/\Rout=0.3$ (see Fig.~\ref{fig:cygx1lag}, we assumed $\beta=0$, $\cosi=0.7$, 
and $\alpha=3$). 
One can see that reflection,  however, cannot reproduce the observed sharp features. 
For  the given parameters, the amplitude of reflection is close to that observed 
in Cyg X-1 (e.g.  Gierli\'nski  et al.  1997; Poutanen  1998).
In order not to overproduce lags at high frequencies, 
the reflection should be suppressed at small distances from the 
central source. This can be achieved by truncation of the disc 
at $\sim 30\Rg$ (as here), by ionisation  of the disc, or  beaming
of the radiation  away from the disc. 
A rather small size of the disc is consistent with the fact that 
Cyg X-1  is probably accreting via wind from the high-mass
companion. The angular momentum of accreting material is then small and the 
disc is small (Illarionov \& Sunyaev 1975; Beloborodov \& Illarionov  2001). 
In low-mass X-ray binaries where  material  accretes via Roche lobe  overflow,
the angular  momentum  is high and the disc is large.  In such  systems the
phase lag secondary maxima should appear at lower frequencies as seems to be
the case (Nowak et al.  1999b;  Grove et al.  1998).  Thus, if the model is
correct, the disc size can be determined from the position of the phase lag
peak.  The lags energy  dependence at low frequencies can resolve the issue
whether the phase lag breaks are due to reflection.

Revnivtsev, Gilfanov \& Churazov (1999) showed that the  frequency-resolved
spectra  (proportional to the amplitude of the Fourier transforms $\hID_E(f)$
at a given  frequency  $f$ as a  function  of  $E$) of Cyg X-1  depends  on
frequency.  At $f>20$  Hz, the  spectra  are  rather  hard and  contain  no
reflection  features.  At low  frequencies  (below 0.1 Hz), the spectra are
much  softer  and show  strong  reflection  features.  If the  response  is
linear, reflection modifies the frequency resolved spectra in the following
way   $|\hI_E(f)|\approx   |\hID_E(f)|   [1+\aE  \Re  \hT(f)]$   (see  also
eqs.~\ref{eq:iefou},  \ref{eq:iepds}).  The straightforward  explanation of
the observed  effects as solely  caused by the finite light  crossing  time
(studied in the present  paper)  fails, since the delays due to  reflection
cannot change the slope of the underlying  spectra.  The observed behaviour
implies  that  the  intrinsic  (Comptonized)   frequency-resolved  spectra,
$\hID_E(f)$,  depend on  frequency.  In that case, the signal at  different
frequencies  possibly is produced in physically  different  locations.  For
example,  the high  frequency  signal can be  produced  closer to the black
hole, while the low frequency signal--further away.  This arrangement would
be natural if the dominant  frequencies  are the Keplerian  ones.  However,
physically  different  locations also imply different  reflection  response
functions thus complicating the situation.

The  X-ray  reprocessing  in the  cool  disc  produces  both  the  Compton
reflection  continuum with an Fe line and soft  radiation.  The reprocessed
radiation  from  the  outer  disc  emerges  in the  optical/UV  bands.  The
response  function  of  this  soft  radiation  to  the  variability  of the
intrinsic  X-ray  radiation would be similar to that of the Compton  reflection
continuum unless ionisation state of the reflecting material is a
function of radius. Thus, the accretion  disc response  function  obtained from the
cross-correlation  analysis of the optical and the X-ray light  curves (see
e.g.  Hynes et al.  1998;  O'Brien \& Horne 2001)  should also  satisfy the
constraints    coming    from   the    X-ray    timing    e.g.   from   the
Fourier-frequency--dependent phase lags.  If the response is known from the
optical/X-ray  data, then one can try to subtract  the lags  related to the
reflection  from  the  observed  time  lags in  order  to  obtain  the lags
corresponding  to the  intrinsic  (direct)  signal  only.  This will not be
easy, however, if the reflection response is non-linear or the disc 
is ionised (then the UV and Compton reflection responses are different).

Real accretion discs are not necessarily  axisymmetric, but can be, for
example, radiatively warped (e.g.  Pringle 1996; Wijers \& Pringle 1999) or
warped due to the  Bardeen--Peterson  (1975) effect.  For  non-axisymmetric
discs the response  function depends not only on the inclination,  but also
on the azimuthal  angle.  Blackman  (1999) and Hartnoll \& Blackman  (2000)
studied the Fe line  profiles  produced in flared  axisymmetric  and warped
discs.  It would be interesting to study temporal response of the iron line
profile for such discs.  If reflection is responsible for shaping the phase
lags, they would change periodically depending on the azimuthal angle.

We focused on the   effects  of  Compton   reflection   on   temporal
characteristics and  not  much  attention  was  paid  to  the  iron
emission line.  With future  instruments  of high spectral  resolution  and
large  effective  area, it will be possible  to study the phase lags across
the line profile.  By combining the line profile with the energy dependence
of the phase lags it will be  possible  to put  better  constraints  on the
accretion disc geometry in X-ray  binaries.  The narrow cores,  produced in
the material far away from the X-ray source, would be associated with large
time delays.  The broad wings should  respond  faster to the  variations of
the  continuum  and the lags should be  smaller.  However,  there  could be
serious  complications.  We  assumed  everywhere  that  there  is a  linear
response between  variations of the continuum and the reflected  radiation.
In other words,  properties  of the  reflecting  material are assumed to be
constant  independently of the X-ray flux.  This assumption  could be wrong
since  the  ionisation  state of the  reflecting  material  depends  on the
illuminating flux (see e.g.  Reynolds 2000; Nayakshin \& Kazanas 2001).

Another  complication  is that the angular  distribution  of the  intrinsic
radiation  could change.  This would be likely for example in the model with
mildly-relativistic  outflows  (Beloborodov  1999; Malzac,  Beloborodov  \&
Poutanen 2001).  If the velocity of the plasma ejection is correlated  with
the energy  dissipation  rate in magnetic  flares, one could  expect that a
large emitted  luminosity  corresponds to a larger  ejection  velocity, and
more beaming  away from the disc.  This would  decrease  the  amplitude  of
reflection at large fluxes.  The resulting  reflection  response could thus
be  non-linear  possibly   reproducing   negative  lags  introduced  by
reflection  (Kotov et al.  2001).  Further studies in that direction are in
progress.

\section{Conclusions} \label{sect:concl}

In the present  paper, we have  developed the formalism  for  computing the
response  functions for isotropic and anisotropic  X-ray sources above flat
and  flared  accretion  discs.  We have  also  studied  the  impact  of the
reflection  on the  temporal  characteristics  such  as the  power  density
spectra,    auto-    and     cross-correlation     functions,    and    the
Fourier-frequency--dependent   time/phase   lags.  A   number   of   useful
approximations  was introduced to simplify  significantly  computations of
the response function.  Simple approximate  response function was shown to
have temporal  characteristics which are very similar to that computed using
exact responses.

The temporal characteristics predicted by a model with a linear response of
the reflected  component onto the  variability  of the intrinsic  radiation
were computed.  The reflection  model can reproduce the observed  secondary
maxima at low  frequencies in the phase lag Fourier  spectra of Cyg X-1 and
other GBHs.  The position of these maxima could be used to get  constraints
on the accretion disc size.  In spite of the fact that the observed  energy
dependence of the phase lags at high  frequencies  indicate that reflection
is not their  only  source  (Kotov et al.  2001), a proper  account  of the
impact of the  reflection on temporal  characteristics  is necessary,  when
interpreting the data.

\section*{Acknowledgments}

This work was supported by the Swedish Natural Science Research Council and
the Anna-Greta and Holger  Crafoord Fund.  The author is grateful to Andrei
Beloborodov and an anonymous referee  
for  useful   comments   that   significantly   improved  the paper.

\begin{figure*}
\centerline{\epsfig{file=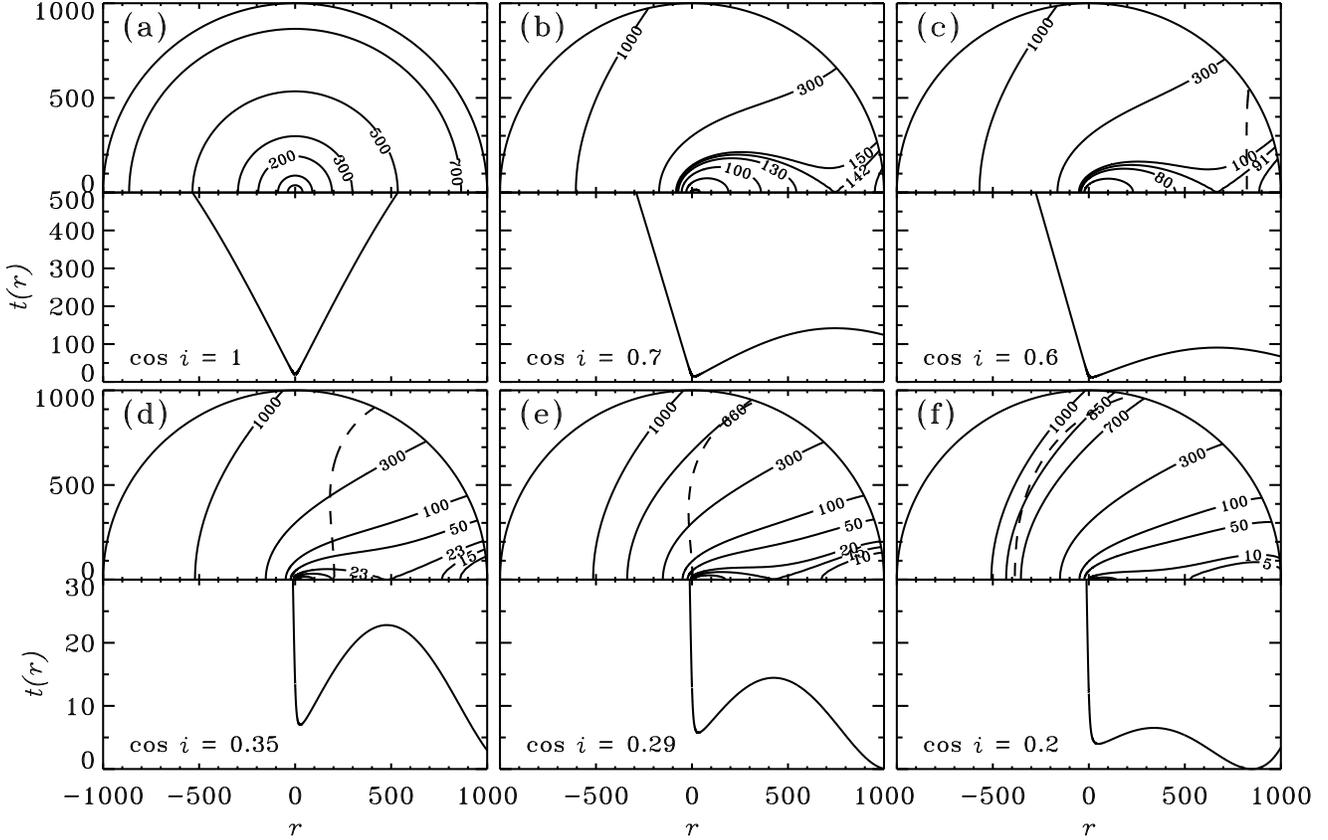,width=17cm}}
\caption{ 
{\it Upper  panels}:  The curves of equal  delays  (isochrons)  at the disc
surface for  different  inclinations  (only one half of the disc is shown).
The   reflector's   surface  is  given  by   equation   (\ref{eq:z})   with
$H/\Rout=0.3, \alpha=3, \Rout=1000, \Rin=0$.  The flare elevation above the
disc center  $h=10$.  The  observer is  situated  in the  $x$-$z$  plane at
inclination  $i$.  For large  inclinations  and  certain  time  delays, the
isochrons consist of two disconnected  regions:  one is close to the flare,
the  other  one is close to the disc  edge  (see e.g.  panels b and c).  At
very  large  inclinations,  a  certain  area on the  disc  surface  becomes
invisible     (rightwards    from    the    dashed    curves    given    by
eq.~\ref{eq:cosphi0}).       
{\it  Lower  panels}:       Function $t(r)=p-(z-h)\cosi-r\sini$.
}
\label{fig:isochrons}
\end{figure*}

\appendix

\section{Calculation of the response function}

\subsection{Visibility conditions}

Contrary to the case of the flat disc,  the every point of a
flared disc is  not  necessarily visible.  For a concave disc surface, $z''>0$, the condition
for visibility reads
\be \label{eq:visib}
\tan i < \frac{\sqrt{\Rout^2-r^2\sin^2\phi}-r\cos\phi}{H-z} .
\ee
For a given inclination $i$ and radius $r$, this translates to
\be  \label{eq:cosphi0}
\cos\phi< \cos\phi_0\equiv \frac{(\Rout^2-r^2)-(H-z)^2 \tan^2i} 
{2(H-z)r\tan i} .
\ee
According to the value of $\tan i$, the  visibility  condition  is:  (i) if
$\tan  i<1/z'_{\max}$,  then the whole disc  surface  is  visible;  (ii) if
$1/z'_{\max}<\tan  i<\Rout/H$,  then the whole area  $r<r_{c}$  is visible.
For   $r_{c}<r<\Rout$, the  condition   $\cos\phi<\cos\phi_0$   should   be
fulfilled.  Here   $r_{c}$  is  the   solution   of  the   equation   $\tan
i=(\Rout-r)/(H-z)$;  (iii) if $\tan  i>\Rout/H$, then the area $r<r_{c}$ is
invisible.  For $r>r_{c}$, the inequality  $\cos\phi<\cos\phi_0$  gives the
visible  area.  Here  $r_{c}$  is  the  solution  of  the  equation   $\tan
i=(\Rout+r)/(H-z)$.

\subsection{Isochrons}

An a given moment of time, the observer receives the signal reflected  from
the curve which is the  interception of the paraboloid of equal delays with
the disc  surface.  For an  infinite  flat disc,  this  curve is an ellipse
with a center shifted along the $x$-axis:
\be
\frac{(x-q\sin^2i\ \cosi)^2}{(s/\sini)^2}+\frac{y^2}{(s\ {\rm ctan}\ i)^2}=1,
\ee
where $q$ and $s$ are given by equation  (\ref{eq:ppm}).  A few  examples of
isochrons for a flared disc are presented in  Fig.~\ref{fig:isochrons}.  At
small  inclinations  and small  delays they are close to those for the flat
disc (see  Fig.~\ref{fig:isochrons}b,c).  The curve of equal  delays on the
disc  surface can  consist of two  disconnected  regions  depending  on the
inclination,  disc  parameters and time.  Some part of the disc can also be
blocked  from the  observer by the disc outer  edge.  The area on the right
side from the dashed curves on Fig.~\ref{fig:isochrons} is invisible.

\subsection{Integration over radius}
\label{sec:app3} 

Computation of the response  function for  flared discs is reduced to one
integral  over the radius (see  eqs.~\ref{eq:resp},  \ref{eq:respiso}).  For
zero   inclination   $i=0$,  the   integration   is   trivial   using   the
$\delta$-function    from    equation~(\ref{eq:Gr}).   For   the   non-zero
inclination,  the integration  limits $\rmin$,  $\rmax$ are given either by
the solution of the equations
\be \label{eq:findr}
t=p-(z-h)\cosi\pm r\sini,
\ee
or by the minimum/maximum disc radii $\Rin$, $\Rout$.  Integrating  over
radius  we use a  substitution  $r=[\rmax+\rmin+(\rmax-\rmin)\cos\theta]/2$
which  removes  possible  divergency  of the  integrand at the  integration
limits.  The solution of  equation~(\ref{eq:findr})  is trivial in the case
of $z=0$  (see  eqs.~\ref{eq:ppm},  \ref{eq:tflat0}).  In  order  to  solve
equation~(\ref{eq:findr})  for $z\neq 0$, we first  tabulate  the  function
$t(r)=p-(z-h)\cosi-r\sini$ at a dense grid of $r$, with $r$ varying between
$-\Rout$ and $\Rout$.  If $\tan i<\Rout/H$, equation  (\ref{eq:findr})  can
have from 1 up to 3  solutions  (for  $-\Rout<r<\Rout$).  If  $t'(\Rout)>0$,
then there are two solutions for $t_{\min}<t\le t(\Rout)$ that are found by
interpolation   ($t_{\min}$   is  the   minimum   value  of   $t(r)$).  For
$t(-\Rout)<t<t(\Rout)$,  $\rmax=\Rout$  and the only  solution  is found by
interpolation  ($\rmin$ is then the absolute  value of that).  The response
function   for   $t\ge   t(-\Rout)$   is   zero.   If   $t(\Rout)<0$   (see
Fig.~\ref{fig:isochrons}b,c),  the  integral  over radius has to be divided
into two parts, since the isochrons  consist of two  separated  regions for
some $t$.

For every $t$, we check that the visibility condition (\ref{eq:cosphi0}) is
satisfied at all  integration  points  $\theta$.  If the  condition  is not
satisfied  for some  points,  we find  the  radius  (or  radii)  satisfying
equation    $\cos\phi(r)    =    \cos\phi_0(r)$    by    iterations    (see
eqs.~\ref{eq:cosphi},  \ref{eq:cosphi0}).  The integral is then  recomputed
only over radii where  (\ref{eq:cosphi0})  valid.  All the routines used in
this paper are written on {\sc IDL}.

\end{document}